\documentclass[aip,jcp,reprint,noshowkeys,superscriptaddress]{revtex4-1}
\usepackage{graphicx,dcolumn,bm,xcolor,microtype,multirow,amscd,amsmath,amssymb,amsfonts,physics,longtable,wrapfig,txfonts}
\usepackage[version=4]{mhchem}

\usepackage[utf8]{inputenc}
\usepackage[T1]{fontenc}
\usepackage{txfonts}

\usepackage[
	colorlinks=true,
    citecolor=blue,
    breaklinks=true
	]{hyperref}
\newcommand{\ie}{\textit{i.e.}}
\newcommand{\eg}{\textit{e.g.}}

\definecolor{darkgreen}{HTML}{009900}
\usepackage[normalem]{ulem}
\newcommand{\titou}[1]{\textcolor{black}{#1}}

\newcommand{\mc}{\multicolumn}
\newcommand{\fnm}{\footnotemark}
\newcommand{\fnt}{\footnotetext}
\newcommand{\tabc}[1]{\multicolumn{1}{c}{#1}}

\newcommand{\T}[1]{#1^{\intercal}}

\newcommand{\br}{\mathbf{r}}

\newcommand{\GOWO}{$G_0W_0$}

\newcommand{\Norb}{N_\text{orb}}
\newcommand{\Nocc}{O}
\newcommand{\Nvir}{V}

\newcommand{\hH}{\Hat{H}}
\newcommand{\ha}{\Hat{a}}

\newcommand{\KS}{\text{KS}}

\newcommand{\RPA}{\text{RPA}}

\newcommand{\GW}{GW}
\newcommand{\stat}{\text{stat}}
\newcommand{\dyn}{\text{dyn}}

\newcommand{\e}[1]{\eps_{#1}}

\newcommand{\eGW}[1]{\eps^{GW}_{#1}}

\newcommand{\Om}[2]{\Omega_{#1}^{#2}}

\newcommand{\homu}{\frac{{\omega}_1}{2}}

\newcommand{\A}[2]{A_{#1}^{#2}}

\newcommand{\B}[2]{B_{#1}^{#2}}

\newcommand{\W}[2]{W_{#1}^{#2}}
\newcommand{\tW}[2]{\widetilde{W}_{#1}^{#2}}

\newcommand{\MO}[1]{\phi_{#1}}
\newcommand{\ERI}[2]{(#1|#2)}
\newcommand{\sERI}[2]{[#1|#2]}

\newcommand{\OmRPA}[1]{\Omega_{#1}^{\text{RPA}}}

\newcommand{\bA}[1]{\mathbf{A}^{#1}}

\newcommand{\bB}[1]{\mathbf{B}^{#1}}
\newcommand{\bX}[2]{\mathbf{X}_{#1}^{#2}}
\newcommand{\bY}[2]{\mathbf{Y}_{#1}^{#2}}

\newcommand{\eps}{\varepsilon}

\newcommand{\HOMO}{\text{HOMO}}
\newcommand{\LUMO}{\text{LUMO}}
\newcommand{\Eg}{E_\text{g}}
\newcommand{\EgFun}{\Eg^\text{fund}}
\newcommand{\EgOpt}{\Eg^\text{opt}}
\newcommand{\EB}{E_B}

\newcommand{\si}{\sigma}

\newcommand{\pis}{\pi^*}
\newcommand{\ra}{\rightarrow}

\newcommand{\hpsi}{\Hat{\psi}}
\renewcommand{\ha}{\Hat{a}}
\newcommand{\tchi}{\Tilde{\chi}}
\newcommand{\bx}{\mathbf{x}}

\newcommand{\NEEL}{Universit\'e Grenoble Alpes, CNRS, Institut NEEL, F-38042 Grenoble, France}
\newcommand{\LCPQ}{Laboratoire de Chimie et Physique Quantiques (UMR 5626), Universit\'e de Toulouse, CNRS, UPS, France}

\begin{document}	

\title{Dynamical Correction to the Bethe-Salpeter Equation Beyond the Plasmon-Pole Approximation}

\author{Pierre-Fran\c{c}ois \surname{Loos}}
	\email{loos@irsamc.ups-tlse.fr}
	\affiliation{\LCPQ}
\author{Xavier \surname{Blase}}
	\email{xavier.blase@neel.cnrs.fr }
	\affiliation{\NEEL}

\begin{abstract}
The Bethe-Salpeter equation (BSE) formalism is a computationally affordable method for the calculation of accurate optical excitation energies in molecular systems.
Similar to the ubiquitous adiabatic approximation of time-dependent density-functional theory, the static approximation, which substitutes a dynamical (\ie, frequency-dependent) kernel by its static limit, is usually enforced in most implementations of the BSE formalism.
Here, going beyond the static approximation, we compute the dynamical correction of the electron-hole screening for molecular excitation energies thanks to a renormalized first-order perturbative correction to the static BSE excitation energies.
The present dynamical correction goes beyond the plasmon-pole approximation as the dynamical screening of the Coulomb interaction is computed exactly within the random-phase approximation.
Our calculations are benchmarked against high-level (coupled-cluster) calculations, allowing to assess the clear improvement brought by the dynamical correction for both singlet and triplet optical transitions.
%\\
%\bigskip
%\begin{center}
%	\boxed{\includegraphics[width=0.5\linewidth]{TOC}}
%\end{center}
%\bigskip
\end{abstract}

\maketitle

%%%%%%%%%%%%%%%%%%%%%%%%
\section{Introduction}
\label{sec:intro}
%%%%%%%%%%%%%%%%%%%%%%%%
The Bethe-Salpeter equation (BSE) formalism \cite{Salpeter_1951,Strinati_1988} is to the $GW$ approximation \cite{Hedin_1965,Golze_2019} of many-body perturbation theory (MBPT) \cite{Onida_2002,Martin_2016} what time-dependent density-functional theory (TD-DFT) \cite{Runge_1984,Casida_1995} is to Kohn-Sham density-functional theory (KS-DFT), \cite{Hohenberg_1964,Kohn_1965} an affordable way of computing the neutral (or optical) excitations of a given electronic system. 
In recent years, it has been shown to be a valuable tool for computational chemists with a large number of systematic benchmark studies on large families of molecular systems appearing in the literature \cite{Boulanger_2014,Jacquemin_2015a,Bruneval_2015,Jacquemin_2015b,Hirose_2015,Jacquemin_2017a,Jacquemin_2017b,Rangel_2017,Krause_2017,Gui_2018,Liu_2020} (see Ref.~\onlinecite{Blase_2018} for a recent review).

Qualitatively, taking the optical gap (\ie, the lowest optical excitation energy) as an example, BSE builds on top of a $GW$ calculation by adding up excitonic effects (\ie, the electron-hole binding energy $\EB$) to the $GW$ HOMO-LUMO gap
\begin{equation}
	\Eg^{\GW} = \eps_{\LUMO}^{\GW} - \eps_{\HOMO}^{\GW},
\end{equation}
which is itself a corrected version of the Kohn-Sham (KS) gap
\begin{equation}
	\Eg^{\KS} = \eps_{\LUMO}^{\KS} - \eps_{\HOMO}^{\KS} \ll \Eg^{\GW} \approx \EgFun,
\end{equation}
in order to approximate the optical gap
\begin{equation}
	\EgOpt = E_1^{N} - E_0^{N} = \EgFun + \EB,
\end{equation}
where 
\begin{equation} \label{eq:Egfun}
	\EgFun = I^N - A^N
\end{equation}
is the fundamental gap, $I^N = E_0^{N-1} - E_0^N$ and $A^N = E_0^{N} - E_0^{N+1}$ being the ionization potential and the electron affinity of the $N$-electron system, respectively.
Here, $E_S^{N}$ is the total energy of the $S$th excited state of the $N$-electron system, and $E_0^N$ corresponds to its ground-state energy.
Because the excitonic effect corresponds physically to the stabilization implied by the attraction of the excited electron and its hole left behind, we have $\EgOpt < \EgFun$.
Due to the smaller amount of screening in molecules as compared to solids, a faithful description of excitonic effects is paramount in molecular systems.

Most of BSE implementations rely on the so-called static approximation, which approximates the dynamical (\ie, frequency-dependent) BSE kernel by its static limit.
In complete analogy with the ubiquitous adiabatic approximation in TD-DFT where the exchange-correlation (xc) kernel is made static, one key consequence of the static approximation within BSE is that double (and higher) excitations are completely absent from the BSE spectrum.
Indeed, a frequency-dependent kernel has the ability to create additional poles in the response function, which describe states with a multiple-excitation character, and, in particular, double excitations.
Although these double excitations are usually experimentally dark (which means that they usually cannot be observed in photo-absorption spectroscopy), these states play, indirectly, a key role in many photochemistry mechanisms, \cite{Boggio-Pasqua_2007} as they strongly mix with the bright singly-excited states leading to the formation of satellite peaks. \cite{Helbig_2011,Elliott_2011}
They are particularly important in the faithful description of the ground state of open-shell molecules, \cite{Casida_2005,Romaniello_2009a,Huix-Rotllant_2011,Loos_2020c}
and they are, moreover, a real challenge for high-level computational methods. \cite{Loos_2018a,Loos_2019,Loos_2020b,Loos_2020c}
Double excitations play also a significant role in the correct location of the excited states of polyenes that are closely related to rhodopsin, a biological pigment found in the rods of the retina and involved in the visual transduction. \cite{Olivucci_2010,Robb_2007,Manathunga_2016}
In butadiene, for example, while the bright $1 ^1B_u$ state has a clear $\HOMO \ra \LUMO$ single-excitation character, the dark $2 ^1A_g$ state includes a substantial fraction of doubly-excited character from the $\HOMO^2 \ra \LUMO^2$ double excitation (roughly $30\%$), yet with dominant contributions from the $\HOMO-1 \ra \LUMO$ and $\HOMO \ra \LUMO+1$ single excitations. \cite{Maitra_2004,Cave_2004,Saha_2006,Watson_2012,Shu_2017,Barca_2018a,Barca_2018b,Loos_2019}

Going beyond the static approximation is difficult and very few groups have been addressing the problem. \cite{Strinati_1988,Rohlfing_2000,Sottile_2003,Myohanen_2008,Ma_2009a,Ma_2009b,Romaniello_2009b,Sangalli_2011,Huix-Rotllant_2011,Sakkinen_2012,Zhang_2013,Rebolini_2016,Olevano_2019,Lettmann_2019}
Nonetheless, it is worth mentioning the seminal work of Strinati \titou{(who originally derived the dynamical correction to the BSE)} on core excitons in semiconductors, \cite{Strinati_1982,Strinati_1984,Strinati_1988} in which the dynamical screening effects were taken into account through the dielectric matrix, and where he observed an increase of the binding energy over its value for static screening and a narrowing of the Auger width below its value for a core hole.
Following Strinati's footsteps, Rohlfing and coworkers have developed an efficient way of taking into account, thanks to first-order perturbation theory, the dynamical effects via a plasmon-pole approximation combined with the Tamm-Dancoff approximation (TDA). \cite{Rohlfing_2000,Ma_2009a,Ma_2009b,Baumeier_2012b}
With such a scheme, they have been able to compute the excited states of biological chromophores, showing that taking into account the electron-hole dynamical screening is important for an accurate description of the lowest $n \ra \pi^*$ excitations. \cite{Ma_2009a,Ma_2009b,Baumeier_2012b}
Indeed, studying PYP, retinal and GFP chromophore models, Ma \textit{et al.}~found that \textit{``the influence of dynamical screening on the excitation energies is about $0.1$ eV for the lowest $\pi \ra \pis$ transitions, but for the lowest $n \ra \pis$ transitions the influence is larger, up to $0.25$ eV.''} \cite{Ma_2009b}
A similar conclusion was reached in Ref.~\onlinecite{Ma_2009a}.
Zhang \textit{et al.}~have studied the frequency-dependent second-order Bethe-Salpeter kernel and they have observed an appreciable improvement over configuration interaction with singles (CIS), time-dependent Hartree-Fock (TDHF), and adiabatic TD-DFT results. \cite{Zhang_2013}
Rebolini and Toulouse have performed a similar investigation in a range-separated context, and they have reported a modest improvement over its static counterpart. \cite{Rebolini_2016,Rebolini_PhD} 
In these two latter studies, they also followed a (non-self-consistent) perturbative approach within the TDA with a renormalization of the first-order perturbative correction.

It is important to note that, although all the studies mentioned above are clearly going beyond the static approximation of BSE, they are not able to recover additional excitations as the perturbative treatment accounts for dynamical effects only on excitations already present in the static limit. However, it does permit to recover, for transitions with a dominant single-excitation character, additional relaxation effects coming from higher excitations.
These higher excitations would be explicitly present in the BSE Hamiltonian by ``unfolding'' the dynamical BSE kernel, and one would recover a linear eigenvalue problem with, nonetheless, a much larger dimension. \cite{Loos_2020f}

Based on a simple two-level model which permits to analytically solve the dynamical equations, Romaniello and coworkers \cite{Romaniello_2009b,Sangalli_2011} evidenced that one can genuinely access additional excitations by solving the non-linear, frequency-dependent eigenvalue problem. 
For this particular system, it was shown that a BSE kernel based on the random-phase approximation (RPA) produces indeed double excitations but also unphysical excitations. \cite{Romaniello_2009b} 
The appearance of these spurious excitations was attributed to the self-screening problem. \cite{Romaniello_2009a}
This was fixed in a follow-up paper by Sangalli \textit{et al.} \cite{Sangalli_2011} thanks to the design of a number-conserving approach based on the folding of the second-RPA Hamiltonian, \cite{Wambach_1988} which includes explicitly both single and double excitations.
By computing the polarizability of two unsaturated hydrocarbon chains, \ce{C8H2} and \ce{C4H6}, they showed that their approach produces the correct number of physical excitations.

Finally, let us mention efforts to borrow ingredients from BSE in order to go beyond the adiabatic approximation of TD-DFT. 
For example, Huix-Rotllant and Casida \cite{Casida_2005,Huix-Rotllant_2011} proposed a nonadiabatic correction to the xc kernel using the formalism of superoperators, which includes as a special case the dressed TD-DFT method of Maitra and coworkers, \cite{Maitra_2004,Cave_2004,Elliott_2011,Maitra_2012} where a frequency-dependent kernel is build \textit{a priori} and manually for a particular excitation.
Following a similar strategy, Romaniello \textit{et al.} \cite{Romaniello_2009b} took advantages of the dynamically-screened Coulomb potential from BSE to obtain a dynamic TD-DFT kernel.
In this regard, MBPT provides key insights about what is missing in adiabatic TD-DFT, as discussed in details by Casida and Huix-Rotllant in Ref.~\onlinecite{Casida_2016}.

In the present study, we extend the work of Rohlfing and coworkers \cite{Rohlfing_2000,Ma_2009a,Ma_2009b,Baumeier_2012b} by proposing a renormalized first-order perturbative correction to the static BSE excitation energies. 
Importantly, our correction goes beyond the plasmon-pole approximation as the dynamical screening of the Coulomb interaction is computed exactly. 
In order to assess the accuracy of the present scheme, we report singlet and triplet excitation energies of various natures for small- and medium-size molecules.
Our calculations are benchmarked against high-level coupled-cluster (CC) calculations, allowing to clearly evidence the systematic improvement brought by the dynamical correction.
In particular, we found that, although $n \ra \pis$ and $\pi \ra \pis$ transitions are systematically red-shifted by $0.3$--$0.6$ eV, dynamical effects have a much smaller  magnitude for charge transfer (CT) and Rydberg states.
Unless otherwise stated, atomic units are used.

%%%%%%%%%%%%%%%%%%%%%%%%
\section{Theory}
\label{sec:theory}
%%%%%%%%%%%%%%%%%%%%%%%%

In this Section, following Strinati's seminal work, \cite{Strinati_1988} we first discuss in some details the theoretical foundations leading to the dynamical BSE. 
%Additional details about this derivation are provided as {\SI}.
We present, in a second step, the perturbative implementation of the dynamical correction as compared to the standard static approximation.

%================================
\subsection{General dynamical BSE}
%=================================

The two-body correlation function $L(1,2; 1',2')$ --- a central quantity in the BSE formalism --- relates the variation of the one-body Green's function $G(1,1')$ with respect to an external non-local perturbation $U(2',2)$, \ie,
\begin{equation}
	iL(1,2; 1',2') = \pdv{G(1,1')}{U(2',2)},
\end{equation}
where, \eg, $1 \equiv (\bx_1 t_1)$ is a space-spin plus time composite variable. 
The relation between $G$ and the one-body charge density $\rho(1) = -i G(1,1^+)$ provides a direct connection with the density-density susceptibility $\chi(1,2) = L(1,2;1^+,2^+)$ at the core of TD-DFT.
(The notation $1^+$ means that the time $t_1$ is taken at $t_1^{+} = t_1 + 0^+$, where $0^+$ is a positive infinitesimal.)

The two-body correlation function $L$ satisfies the self-consistent BSE \cite{Strinati_1988}  
\begin{multline} \label{eq:BSE}
	L(1,2; 1',2') = L_0(1,2;1',2')
	\\
	+ \int d3456 \, L_0(1,4;1',3) \Xi(3,5;4,6) L(6,2;5,2'),
\end{multline}
where
\begin{subequations}
\begin{align}
	\label{eq:L0}
	iL_0(1, 4; 1', 3) & = G(1, 3)G(4, 1'),
	\\
	\label{eq:L}
	iL(1,2; 1',2') & = - G_2(1,2;1',2') + G(1,1') G(2,2'),
\end{align}
\end{subequations}
can be expressed as a function of the one- and two-body Green's functions
\begin{subequations}
\begin{align}
	\label{eq:G1}
	G(1,2) & = - i \mel{N}{T [ \hpsi(1) \hpsi^{\dagger}(2) ] }{N},
	\\
	\label{eq:G2}
	G_2(1,2;1',2') & = - \mel{N}{T [ \hpsi(1) \hpsi(2) \hpsi^{\dagger}(2') \hpsi^{\dagger}(1') ]}{N},
\end{align}
\end{subequations}
and
\begin{equation}
	\Xi(3,5;4,6) = i \fdv{[v_\text{H}(3) \delta(3,4) + \Sigma_\text{xc}(3,4)]}{G(6,5)}
\end{equation}
is the BSE kernel that takes into account the self-consistent variation of the Hartree potential 
\begin{equation}
	v_\text{H}(1) = - i \int d2 \, v(1,2) G(2,2^+),
\end{equation}
[where $\delta$ is Dirac's delta function and $v$ is the bare Coulomb operator] and the xc self-energy $ \Sigma_\text{xc}$ with respect to the variation of $G$.  
In Eqs.~\eqref{eq:G1} and \eqref{eq:G2}, the field operators $\Hat{\psi}(\bx t)$ and $\Hat{\psi}^{\dagger}(\bx't')$ remove and add (respectively) an electron to the $N$-electron ground state $\ket{N}$ in space-spin-time positions ($\bx t$) and ($\bx't'$), while $T$ is the time-ordering operator.

The resolution of the dynamical BSE starts with the expansion of $L_0$ and $L$ [see Eqs.~\eqref{eq:L0} and \eqref{eq:L}] over the complete orthonormalized set of $N$-electron excited states $\ket{N,S}$ (with $\ket{N,0} \equiv \ket{N}$). \cite{Strinati_1988} 
In the optical limit of instantaneous electron-hole creation and destruction, imposing $t_{2'} = t_2^+$ and $t_{1'} = t_1^+$, and using the relation between the field operators in their time-dependent (Heisenberg) and time-independent (Schr\"{o}dinger) representations, \eg,
\begin{equation} \label{Eisenberg}
	\hpsi(1) = e^{ i \hH t_1 } \hpsi(\bx_1) e^{-i \hH t_1 },
\end{equation}
($\hH$ being the exact many-body Hamiltonian), one gets
\begin{equation}
\begin{split}
	iL(1,2; 1',2') 
	& = \theta(+\tau_{12}) \sum_{s > 0} \chi_S(\bx_1,\bx_{1'}) \tchi_S(\bx_2,\bx_{2'}) e^{ - i \Om{S}{} \tau_{12} } 
 	\\
	& - \theta(-\tau_{12}) \sum_{s > 0} \chi_S(\bx_2,\bx_{2'}) \tchi_S(\bx_1,\bx_{1'}) e^{ + i \Om{S}{} \tau_{12} },
\end{split}
\end{equation}
where $\tau_{12} = t_1 - t_2$, $\theta$ is the Heaviside step function, and
\begin{subequations}
\begin{align}
	\chi_S(\bx_1,\bx_{1'}) & = \mel{N}{T [\hpsi(\bx_1) \hpsi^{\dagger}(\bx_{1'})] }{N,S},
	\\
	\tchi_S(\bx_1,\bx_{1'}) & = \mel{N,S}{T [\hpsi(\bx_1) \hpsi^{\dagger}(\bx_{1'})] }{N}.
\end{align}  
\end{subequations}  
The $\Om{s}{}$'s are the neutral excitation energies of interest (with $\Om{s}{} = E^N_s - E^N_0$). 

Picking up the $e^{+i \Om{S}{} t_2 }$ component of both $L(1,2; 1',2')$ and $L(6,2;5,2')$, simplifying further by $\tchi_S(\bx_2,\bx_{2'})$ on both sides of the BSE [see Eq.~\eqref{eq:BSE}], we seek the $e^{-i \Om{S}{} t_1 }$ Fourier component associated with the right-hand side of a modified dynamical BSE, which reads
\begin{multline} \label{eq:BSE_2}
    \mel{N}{T [ \hpsi(\bx_1) \hpsi^{\dagger}(\bx_{1}') ] } {N,S} e^{ - i \Om{S}{} t_1 }
	\theta ( \tau_{12} )  
    \\
	= \int d3456 \, L_0(1,4;1',3) \Xi(3,5;4,6)  
    \\
	\times \mel{N}{T [\hpsi(6) \hpsi^{\dagger}(5)] }{N,S} 
	\theta [\min(t_5,t_6) - t_2].
\end{multline}
For the neutral excitation energies falling in the fundamental gap of the system (\ie, $\Om{S}{} < \EgFun$ due to excitonic effects), $L_0(1,2;1',2')$ cannot contribute to the $e^{-i \Om{S}{} t_1 }$ response  term since its lowest excitation energy is precisely the fundamental gap [see Eq.~\eqref{eq:Egfun}].
Consequently, special care has to be taken for high-lying excited states (like core or Rydberg excitations) where additional terms have to be taken into account (see Refs.~\onlinecite{Strinati_1982,Strinati_1984}).

Dropping the space/spin variables, the Fourier components with respect to $t_1$ of $L_0(1,4;1',3)$ reads
\begin{align} \label{eq:iL0}
	[iL_0]( \omega_1 ) 
	= \int \frac{d\omega}{2\pi} \; G\qty(\omega - \frac{\omega_1}{2} ) G\qty( {\omega} + \frac{\omega_1}{2} )
	e^{ i \omega \tau_{34} } e^{ i \omega_1 t^{34} },
   \end{align}
with $\tau_{34} = t_3 - t_4$ and $t^{34} = (t_3 + t_4)/2$. 
We now adopt the Lehman representation of the one-body Green's function in the quasiparticle approximation, \ie,
\begin{equation} \label{eq:G-Lehman}
	G(\bx_1,\bx_2 ; \omega) = \sum_p \frac{ \MO{p}(\bx_1) \MO{p}^*(\bx_2) } { \omega - \e{p} + i \eta \times \text{sgn}  (\e{p} - \mu) },
\end{equation}
where $\eta$ is a positive infinitesimal and $\mu$ is the chemical potential. 
The $\e{p}$'s in Eq.~\eqref{eq:G-Lehman} are quasiparticle energies (\ie, proper addition/removal energies) and the $\MO{p}(\bx)$'s are their associated one-body (spin)orbitals.
In the following, $i$ and $j$ are occupied orbitals, $a$ and $b$ are unoccupied orbitals, while $p$, $q$, $r$, and $s$ indicate arbitrary orbitals.
Projecting the Fourier component $L_0(\bx_1,4;\bx_{1'},3;  \omega_1 = \Om{S}{} )$ onto $\MO{a}^*(\bx_1) \MO{i}(\bx_{1'})$ yields
\begin{multline}  \label{eq:iL0bis} 
	\iint d\bx_1 d\bx_{1'} \, \MO{a}^*(\bx_1) \MO{i}(\bx_{1'})  L_0(\bx_1,4;\bx_{1'},3; \Om{S}{}) 
	\\
	= 
	\frac{ \MO{a}^*(\bx_3) \MO{i}(\bx_4)  e^{i \Om{S}{} t^{34} }} { \Om{S}{} - ( \e{a} - \e{i} ) + i \eta }   
	\qty[ \theta( \tau_{34} ) e^{i \qty( \e{i} + \frac{\Om{S}{}}{2}) \tau_{34} } + \theta( - \tau_{34} ) e^{i \qty(\e{a} - \frac{\Om{S}{}}{2}) \tau_{34} }  ]. 
\end{multline}
More details are provided in Appendix \ref{app:A}.
As a final step, we express the terms $\mel{N}{T [\hpsi(\bx_1) \hpsi^{\dagger}(\bx_{1}')] }{N,S}$ and $\mel{N}{T [\hpsi(6) \hpsi^{\dagger}(5)] }{N,S}$ from Eq.~\eqref{eq:BSE_2} in the standard electron-hole product (or single-excitation) space.
This is done by  expanding the field operators over a complete orbital basis of creation/destruction operators.
For example, we have (see derivation in Appendix \ref{app:B})
\begin{multline} \label{eq:spectral65}
	\mel{N}{T [\hpsi(6) \hpsi^{\dagger}(5)] }{N,S} 
	\\
	= - \qty( e^{  -i \Om{S}{} t^{65} }  ) \sum_{pq} \MO{p}(\bx_6) \MO{q}^*(\bx_5) 
	\mel{N}{\ha_q^{\dagger} \ha_p}{N,S}
	\\
	\times \qty[ \theta( \tau_{65}  ) e^{- i \qty( \e{p} - \frac{\Om{S}{}}{2} ) \tau_{65}  } + \theta( - \tau_{65}  ) e^{ - i \qty( \e{q}  + \frac{\Om{S}{}}{2}) \tau_{65}  } ], 
\end{multline}
with $t^{65} = (t_5 + t_6)/2$ and $\tau_{65} = t_6 -t_5$. The $\mel{N}{\ha_q^{\dagger} \ha_p}{N,S}$ are the unknown particle-hole amplitudes.% in the orbital product basis.

%================================
\subsection{Dynamical BSE within the $GW$ approximation}
%=================================

Adopting now the $GW$ approximation \cite{Hedin_1965} for the xc self-energy, \ie, 
\begin{equation}
	\Sigma_\text{xc}^{\GW}(1,2) = i G(1,2) W(1^+,2),
\end{equation}
leads to the following simplified BSE kernel
\begin{equation} \label{eq:Xi_GW}
	\Xi(3,5;4,6) = v(3,6) \delta(3,4) \delta(5,6) - W(3^+,4) \delta(3,6) \delta(4,5),
\end{equation}
where $W$ is the dynamically-screened Coulomb operator.
The $GW$ quasiparticle energies $\eGW{p}$ are usually good approximations to the removal/addition energies $\e{p}$ introduced in Eq.~\eqref{eq:G-Lehman}. 

Substituting Eqs.~\eqref{eq:iL0bis}, \eqref{eq:spectral65}, and \eqref{eq:Xi_GW} into Eq.~\eqref{eq:BSE_2},  and projecting onto $\MO{a}^*(\bx_1) \MO{i}(\bx_{1'})$, one gets after a few tedious manipulations the dynamical BSE:
\begin{equation} \label{eq:BSE-final}
\begin{split}
	( \eGW{a} - \eGW{i} - \Om{S}{} ) X_{ia,S}      
	& + \sum_{jb} \qty[ \kappa \ERI{ia}{jb} - \widetilde{W}_{ij,ab}(\Om{S}{}) ] X_{jb,S}  \\ 
	& + \sum_{jb} \qty[ \kappa \ERI{ia}{bj} - \widetilde{W}_{ib,aj}(\Om{S}{}) ] Y_{jb,S} 
	= 0,
\end{split}
\end{equation}
with $X_{jb,S} = \mel{N}{\ha_j^{\dagger} \ha_b}{N,S}$ and $Y_{jb,S} = \mel{N}{\ha_b^{\dagger} \ha_j}{N,S}$, and where $\kappa = 2 $ or $0$ for singlet and triplet excited states (respectively).
\titou{This equation is identical to the one presented by Rohlfing and coworkers. \cite{Rohlfing_2000,Ma_2009a,Ma_2009b}}
Neglecting the anti-resonant terms, $Y_{jb,S}$, in the dynamical BSE, which are (usually) much smaller than their resonant counterparts, $X_{jb,S}$, leads to the well-known TDA. 
In Eq.~\eqref{eq:BSE-final}, 
\begin{equation}
	\ERI{pq}{rs} = \iint d\br d\br' \, \MO{p}(\br) \MO{q}(\br) v(\br -\br') \MO{r}(\br') \MO{s}(\br'),
\end{equation}
are the bare two-electron integrals in the (real-valued) spatial orbital basis $\lbrace \MO{p}(\br{}) \rbrace$, and
\begin{multline} \label{eq:wtilde}
	\widetilde{W}_{pq,rs}(\Om{S}{})
	= \frac{ i }{ 2  \pi}  \int d\omega \; e^{-i \omega 0^+ } W_{pq,rs}(\omega) 
	\\
	\times \qty[ \frac{1}{ \Om{ps}{S} - \omega + i \eta } + \frac{1}{ \Om{qr}{S} + \omega + i\eta } ],
\end{multline}
is an effective dynamically-screened Coulomb potential, \cite{Romaniello_2009b} where $\Om{pq}{S} = \Om{S}{} - ( \eGW{q} - \eGW{p} )$ and
\begin{equation}
	W_{pq,rs}({\omega}) 
	 = \iint d\br d\br' \, \MO{p}(\br) \MO{q}(\br) W(\br ,\br'; \omega) \MO{r}(\br') \MO{s}(\br').
\end{equation}

%\xavier{\sout{ A second coupled equation for the $(X_{ia}^{S}, Y_{ia}^{S} )$ vector can be obtained by projecting $\mel{N}{T [ \hpsi(\bx_1) \hpsi^{\dagger}(\bx_{1}') ] } {N,S}$ and $L_0(\bx_1,4;\bx_{1'},3; \Om{S}{})$ onto $\MO{i}^*(\bx_1) \MO{a}(\bx_{1'})$ instead of $\MO{a}^*(\bx_1) \MO{i}(\bx_{1'})$. } }

%================================
\subsection{Dynamical screening}
\label{sec:dynW}
%=================================

In the present study, we consider the exact spectral representation of $W$ at the RPA level \titou{consistently with the underlying $GW$ calculation}:
\begin{multline}
\label{eq:W-RPA}
	W_{ij,ab}(\omega) 
	= \ERI{ij}{ab} + 2 \sum_m \sERI{ij}{m} \sERI{ab}{m} 
	\\
	\times \qty[  \frac{1}{ \omega-\Om{m}{\RPA} + i\eta } - \frac{1}{ \omega + \Om{m}{\RPA} - i\eta } ],
\end{multline}
where $m$ labels single excitations, and
\begin{equation}
\label{eq:sERI}
	\sERI{pq}{m} =  \sum_{ia} \ERI{pq}{ia} (\bX{m}{\RPA} + \bY{m}{\RPA})_{ia}
\end{equation}
are the spectral weights.
In Eqs.~\eqref{eq:W-RPA} and \eqref{eq:sERI}, $\OmRPA{m}{}$ and $(\bX{m}{\RPA} + \bY{m}{\RPA})$ are RPA neutral excitations and their corresponding transition vectors computed by solving the (static) linear response problem 
\begin{equation}
\label{eq:LR-RPA}
	\begin{pmatrix}
		\bA{\RPA}	&	\bB{\RPA}	\\
		-\bB{\RPA}	&	-\bA{\RPA}	\\
	\end{pmatrix}
	\cdot
	\begin{pmatrix}
		\bX{m}{\RPA}	\\
		\bY{m}{\RPA}	\\
	\end{pmatrix}
	=
	\OmRPA{m}
	\begin{pmatrix}
		\bX{m}{\RPA}	\\
		\bY{m}{\RPA}	\\
	\end{pmatrix},
\end{equation}
with
\begin{subequations}
\begin{align}
	\label{eq:LR_RPA-A}
	\A{ia,jb}{\RPA} & = \delta_{ij} \delta_{ab} (\e{a} - \e{i}) +   2 \ERI{ia}{jb},
	\\ 
	\label{eq:LR_RPA-B}
	\B{ia,jb}{\RPA} & =   2 \ERI{ia}{bj},
\end{align}
\end{subequations}
where the $\e{p}$'s are taken as the HF orbital energies in the case of $G_0W_0$ \cite{Hybertsen_1985a, Hybertsen_1986} or as the $GW$ quasiparticle energies in the case of self-consistent schemes such as ev$GW$. \cite{Hybertsen_1986,Shishkin_2007,Blase_2011,Faber_2011,Rangel_2016,Kaplan_2016,Gui_2018} 
The RPA matrices $\bA{\RPA}$ and $\bB{\RPA}$ in Eq.~\eqref{eq:LR-RPA} are of size $\Nocc \Nvir \times \Nocc \Nvir$, where $\Nocc$ and $\Nvir$ are the number of occupied and virtual orbitals (\ie, $\Norb = \Nocc + \Nvir$ is the total number of spatial orbitals), respectively, and $\bX{m}{\RPA}$, and $\bY{m}{\RPA}$ are (eigen)vectors of length $\Nocc \Nvir$. 

The analysis of the poles of the integrand in Eq.~\eqref{eq:wtilde} yields
\begin{multline} 
	\widetilde{W}_{ij,ab}( \Om{S}{} ) 
	= \ERI{ij}{ab} + 2 \sum_m \sERI{ij}{m} \sERI{ab}{m}  
	\\ 
    \times \qty[ \frac{1}{\Om{ib}{S} - \Om{m}{\RPA} + i\eta} + \frac{1}{\Om{ja}{S} - \Om{m}{\RPA} + i\eta} ].
\end{multline} 
One can verify that, in the static limit where $\Om{m}{\RPA} \to \infty$, the matrix elements $\widetilde{W}_{ij,ab}$ correctly reduce to their static expression
\begin{equation}
\label{eq:Wstat}
	\W{ij,ab}{\text{stat}} 
	\equiv W_{ij,ab}(\omega = 0) 
	= \ERI{ij}{ab} -  4 \sum_m \frac{\sERI{ij}{m} \sERI{ab}{m}}{\OmRPA{m}{} },
\end{equation}
evidencing that the standard static BSE problem is recovered from the present dynamical formalism in this limit.

Due to excitonic effects, the lowest BSE excitation energy, $\Om{1}{}$, stands lower than the lowest RPA excitation energy, $\Om{1}{\RPA}$, so that, $\Om{ib}{S} - \Om{m}{\RPA} < 0 $ and $\widetilde{W}_{ij,ab}(\Om{S}{})$ has no resonances. 
This property holds for low-lying excitations but special care must be taken for higher ones.
Furthermore, $\Om{ib}{S}$ and $\Om{ja}{S}$ are necessarily negative quantities for in-gap low-lying BSE excitations. 
Thus, we have $\abs*{\Om{ib}{S} - \Om{m}{\RPA}}  >  \Om{m}{\RPA}$.
As a consequence, we observe a reduction of the electron-hole screening, \ie, an enhancement of electron-hole binding energy, as compared to the standard static BSE, and consequently smaller (red-shifted) excitation energies. 
This will be numerically illustrated in Sec.~\ref{sec:resdis}.

%================================================
\subsection{Dynamical Tamm-Dancoff approximation}
%================================================
The analysis of the (off-diagonal) screened Coulomb potential matrix elements multiplying the $Y_{jb,S}$ coefficients in Eq.~\eqref{eq:BSE-final}, \ie,
\begin{multline} \label{eq:W-Y}
	\widetilde{W}_{ib,aj}(\Om{S}{}) 
	= \ERI{ib}{aj} + 2 \sum_m \sERI{ib}{m} \sERI{aj}{m}  
	\\ 
    \times \qty[ \frac{1}{\Om{ij}{S} - \Om{m}{\RPA} + i\eta} + \frac{1}{\Om{ba}{S} - \Om{m}{\RPA} + i\eta} ],
\end{multline} 
reveals strong divergences even for low-lying excitations when, for example, $\Om{ba}{S} - \Om{m}{\RPA} = \Om{S}{} - \Om{m}{\RPA} - ( \eGW{a} - \eGW{b} ) \approx 0$.
Such divergences may explain that, in previous studies, dynamical effects were only accounted for at the TDA level. \cite{Strinati_1988,Rohlfing_2000,Ma_2009a,Ma_2009b,Romaniello_2009b,Sangalli_2011,Zhang_2013,Rebolini_2016}
To avoid confusions here, enforcing the TDA for the dynamical correction (which corresponds to neglecting the dynamical correction originating from the anti-resonant part of the BSE Hamiltonian) will be labeled as dTDA in the following.
Going beyond the dTDA is outside the scope of the present study but shall be addressed eventually.

%=================================
\subsection{Perturbative dynamical correction}
%=================================

From a more practical point of view, Eq.~\eqref{eq:BSE-final} can be recast as an non-linear eigenvalue problem and, to compute the BSE excitation energies of a closed-shell system, one must solve the following dynamical (\ie, frequency-dependent) response problem \cite{Strinati_1988}
\begin{equation}
\label{eq:LR-dyn}
	\begin{pmatrix}
		\bA{}(\Om{S}{})		&	\bB{}(\Om{S}{})	
		\\
		-\bB{}(-\Om{S}{})	&	-\bA{}(-\Om{S}{})	
		\\
	\end{pmatrix}
	\cdot
	\begin{pmatrix}
		\bX{S}{}	\\
		\bY{S}{}	\\
	\end{pmatrix}
	=
	\Om{S}{}
	\begin{pmatrix}
		\bX{S}{}	\\
		\bY{S}{}	\\
	\end{pmatrix},
\end{equation}
where the dynamical matrices $\bA{}$ and $\bB{}$ have the same $\Nocc \Nvir \times \Nocc \Nvir$ size than their RPA counterparts, and we assume real quantities from hereon.
Same comment applies to the eigenvectors $\bX{S}{}$, and $\bY{S}{}$ of length $\Nocc \Nvir$.
Note that, due to its non-linear nature, Eq.~\eqref{eq:LR-dyn} may provide more than one solution for each value of $S$. \cite{Romaniello_2009b,Sangalli_2011,Martin_2016}

Accordingly to Eq.~\eqref{eq:BSE-final}, the BSE matrix elements in Eq.~\eqref{eq:LR-dyn} read
\begin{subequations}
\begin{align}
	\label{eq:BSE-Adyn}
	\A{ia,jb}{}(\Om{S}{}) & = \delta_{ij} \delta_{ab} (\eGW{a} - \eGW{i}) + \kappa \ERI{ia}{jb} - \tW{ij,ab}{}(\Om{S}{}),
	\\ 
	\label{eq:BSE-Bdyn}
	\B{ia,jb}{}(\Om{S}{}) & =  \kappa \ERI{ia}{bj} - \tW{ib,aj}{}(\Om{S}{}).
\end{align}
\end{subequations}

Now, let us decompose, using basic Rayleigh-Schr\"odinger perturbation theory, the non-linear eigenproblem \eqref{eq:LR-dyn} as a zeroth-order static (hence linear) reference and a first-order dynamic (hence non-linear) perturbation, such that
\begin{multline}
\label{eq:LR-PT}
	\begin{pmatrix}
		\bA{}(\Om{S}{})		&	\bB{}(\Om{S}{})	\\
		-\bB{}(-\Om{S}{})	&	-\bA{}(-\Om{S}{})	\\
	\end{pmatrix}
	\\
	=
	\begin{pmatrix}
		\bA{(0)}	&	\bB{(0)}	
		\\
		-\bB{(0)}	&	-\bA{(0)}	
		\\
	\end{pmatrix}
	+
	\begin{pmatrix}
		\bA{(1)}(\Om{S}{})		&	\bB{(1)}(\Om{S}{})	\\
		-\bB{(1)}(-\Om{S}{})	&	-\bA{(1)}(-\Om{S}{})	\\
	\end{pmatrix},
\end{multline}
with
\begin{subequations}
\begin{align}
	\label{eq:BSE-A0}
	\A{ia,jb}{(0)} & = \delta_{ij} \delta_{ab} (\eGW{a} - \eGW{i}) + \kappa \ERI{ia}{jb} - \W{ij,ab}{\text{stat}},
	\\ 
	\label{eq:BSE-B0}
	\B{ia,jb}{(0)} & =  \kappa \ERI{ia}{bj} - \W{ib,aj}{\text{stat}}.
\end{align}
\end{subequations}
and 
\begin{subequations}
\begin{align}
	\label{eq:BSE-A1}
	\A{ia,jb}{(1)}(\Om{S}{}) & = - \tW{ij,ab}{}(\Om{S}{}) + \W{ij,ab}{\text{stat}},
	\\ 
	\label{eq:BSE-B1}
	\B{ia,jb}{(1)}(\Om{S}{}) & = -  \tW{ib,aj}{}(\Om{S}{}) + \W{ib,aj}{\text{stat}}.
\end{align}
\end{subequations}

According to perturbation theory, the $S$th BSE excitation energy and its corresponding eigenvector can then be expanded as
\begin{subequations}
\begin{gather}
	\Om{S}{} = \Om{S}{(0)} + \Om{S}{(1)} + \ldots,
	\\
	\begin{pmatrix}
		\bX{S}{}	\\
		\bY{S}{}	\\
	\end{pmatrix}
	= 
	\begin{pmatrix}
		\bX{S}{(0)}	\\
		\bY{S}{(0)}	\\
	\end{pmatrix}
	+
	\begin{pmatrix}
		\bX{S}{(1)}	\\
		\bY{S}{(1)}	\\
	\end{pmatrix}
	+ \ldots.
\end{gather}
\end{subequations}
Solving the zeroth-order static problem
\begin{equation}
\label{eq:LR-BSE-stat}
	\begin{pmatrix}
		\bA{(0)}	&	\bB{(0)}	\\
		-\bB{(0)}	&	-\bA{(0)}	\\
	\end{pmatrix}
	\cdot
	\begin{pmatrix}
		\bX{S}{(0)}	\\
		\bY{S}{(0)}	\\
	\end{pmatrix}
	=
	\Om{S}{(0)}
	\begin{pmatrix}
		\bX{S}{(0)}	\\
		\bY{S}{(0)}	\\
	\end{pmatrix},
\end{equation}
yields the zeroth-order (static) $\Om{S}{(0)}$ excitation energies and their corresponding eigenvectors $\bX{S}{(0)}$ and $\bY{S}{(0)}$. 
Thanks to first-order perturbation theory, the first-order correction to the $S$th excitation energy is
\begin{equation}
\label{eq:Om1}
	\Om{S}{(1)} =
	\T{\begin{pmatrix}
		\bX{S}{(0)}	\\
		\bY{S}{(0)}	\\
	\end{pmatrix}}
	\cdot
	\begin{pmatrix}
		\bA{(1)}(\Om{S}{(0)})	&	\bB{(1)}(\Om{S}{(0)})	\\
		-\bB{(1)}(-\Om{S}{(0)})	&	-\bA{(1)}(-\Om{S}{(0)})	\\
	\end{pmatrix}
	\cdot
	\begin{pmatrix}
		\bX{S}{(0)}	\\
		\bY{S}{(0)}	\\
	\end{pmatrix}.
\end{equation}
From a practical point of view, if one enforces the dTDA, we obtain the very simple expression
\begin{equation}
\label{eq:Om1-TDA}
	\Om{S}{(1)} = \T{(\bX{S}{(0)})} \cdot \bA{(1)}(\Om{S}{(0)}) \cdot \bX{S}{(0)}.
\end{equation}
This correction can be renormalized by computing, at basically no extra cost, the renormalization factor which reads, in the dTDA,
\begin{equation}
\label{eq:Z}
	Z_{S} = \qty[ 1 - \T{(\bX{S}{(0)})} \cdot \left. \pdv{\bA{(1)}(\Om{S}{})}{\Om{S}{}} \right|_{\Om{S}{} = \Om{S}{(0)}} \cdot \bX{S}{(0)} ]^{-1}.
\end{equation}
This finally yields
\begin{equation}
	\Om{S}{\text{dyn}} = \Om{S}{\text{stat}} + \Delta\Om{S}{\text{dyn}} = \Om{S}{(0)} + Z_{S} \Om{S}{(1)}.
\end{equation}
with $\Om{S}{\text{stat}} \equiv \Om{S}{(0)}$ and $\Delta\Om{S}{\text{dyn}} = Z_{S} \Om{S}{(1)}$.
This is our final expression.
\titou{As mentioned in Sec.~\ref{sec:intro}, the present perturbative scheme does not allow to access double excitations as only excitations calculated within the static approach can be dynamically corrected.
We hope to report a genuine dynamical treatment of the BSE in a forthcoming work.}

In terms of computational cost, if one decides to compute the dynamical correction of the $M$ lowest excitation energies, one must perform, first, a conventional (static) BSE calculation and extract the $M$ lowest eigenvalues and their corresponding eigenvectors [see Eq.~\eqref{eq:LR-BSE-stat}].
These are then used to compute the first-order correction from Eq.~\eqref{eq:Om1-TDA}, which also require to construct and evaluate the dynamical part of the BSE Hamiltonian for each excitation one wants to dynamically correct. 
The static BSE Hamiltonian is computed once during the static BSE calculation and does not dependent on the targeted excitation.

Searching iteratively for the lowest eigenstates, via Davidson's algorithm for instance, can be performed in $\order*{\Norb^4}$ computational cost.
Constructing the static and dynamic BSE Hamiltonians is much more expensive as it requires the complete diagonalization of the $(\Nocc \Nvir \times \Nocc \Nvir)$ RPA linear response matrix [see Eq.~\eqref{eq:LR-RPA}], which corresponds to a $\order*{\Nocc^3 \Nvir^3} = \order*{\Norb^6}$ computational cost. 
Although it might be reduced to $\order*{\Norb^4}$ operations with standard resolution-of-the-identity techniques, \cite{Duchemin_2019,Duchemin_2020} this step is the computational bottleneck in the current implementation.

%%%%%%%%%%%%%%%%%%%%%%%%
\section{Computational details}
\label{sec:compdet}
%%%%%%%%%%%%%%%%%%%%%%%%
All systems under investigation have a closed-shell singlet ground state.
We then adopt a restricted formalism throughout this work.
The $GW$ calculations performed to obtain the screened Coulomb operator and the quasiparticle energies are done using a (restricted) HF starting point.
Perturbative $GW$ (or {\GOWO}) \cite{Hybertsen_1985a,Hybertsen_1986,vanSetten_2013} quasiparticle energies are employed as starting points to compute the BSE neutral excitations.
These quasiparticle energies are obtained by linearizing the frequency-dependent quasiparticle equation, and the entire set of orbitals is corrected.
Further details about our implementation of {\GOWO} can be found in Refs.~\onlinecite{Loos_2018b,Veril_2018}.
Note that, for the present (small) molecular systems, {\GOWO}@HF and ev$GW$@HF yield similar quasiparticle energies and fundamental gap. 
Moreover, {\GOWO} allows to avoid rather laborious iterations as well as the significant additional computational effort of ev$GW$.
In the present study, the zeroth-order Hamiltonian [see Eq.~\eqref{eq:LR-PT}] is always the ``full'' BSE static Hamiltonian, \ie, without TDA.
The dynamical correction, however, is computed in the dTDA throughout.
As one-electron basis sets, we employ the Dunning families cc-pVXZ and aug-cc-pVXZ (X = D, T, and Q) defined with cartesian Gaussian functions. 
Finally, the infinitesimal $\eta$ is set to $100$ meV for all calculations.
It is important to mention that the small molecular systems considered here are particularly challenging for the BSE formalism, \cite{Hirose_2015,Loos_2018b} which is known to work best for larger systems where the amount of screening is more important. \cite{Jacquemin_2017b,Rangel_2017}

For comparison purposes, we employ the theoretical best estimates (TBEs) and geometries of Refs.~\onlinecite{Loos_2018a,Loos_2019,Loos_2020b} from which CIS(D), \cite{Head-Gordon_1994,Head-Gordon_1995} ADC(2), \cite{Trofimov_1997,Dreuw_2015} CC2, \cite{Christiansen_1995a} CCSD, \cite{Purvis_1982} and CC3 \cite{Christiansen_1995b} excitation energies are also extracted. 
Various statistical quantities are reported in the following: the mean signed error (MSE), mean absolute error (MAE), root-mean-square error (RMSE), and the maximum positive [Max($+$)] and maximum negative [Max($-$)] errors.
All the static and dynamic BSE calculations have been performed with the software \texttt{QuAcK}, \cite{QuAcK} freely available on \texttt{github}, where the present perturbative correction has been implemented.

%%%%%%%%%%%%%%%%%%%%%%%%
\section{Results and Discussion}
\label{sec:resdis}
%%%%%%%%%%%%%%%%%%%%%%%%

%%% TABLE I %%%
\begin{squeezetable}
\begin{table*}
	\caption{
	Singlet and triplet excitation energies (in eV) of \ce{N2} computed at the BSE@{\GOWO}@HF level for various basis sets.
	\label{tab:N2}
	}
	\begin{ruledtabular}
		\begin{tabular}{lcddddddddd}
						&					&	\mc{3}{c}{cc-pVDZ ($\Eg^{\GW} = 20.71$ eV)}	
											&	\mc{3}{c}{cc-pVTZ ($\Eg^{\GW} = 20.21$ eV)}
											&	\mc{3}{c}{cc-pVQZ ($\Eg^{\GW} = 20.05$ eV)}		\\
											\cline{3-5} \cline{6-8} \cline{9-11}
			State		&	Nature			&	\tabc{$\Om{S}{\stat}$}	&	\tabc{$\Om{S}{\dyn}$}	&	\tabc{$\Delta\Om{S}{\dyn}$} 	
											&	\tabc{$\Om{S}{\stat}$}	&	\tabc{$\Om{S}{\dyn}$}	&	\tabc{$\Delta\Om{S}{\dyn}$}	
											&	\tabc{$\Om{S}{\stat}$}	&	\tabc{$\Om{S}{\dyn}$}	&	\tabc{$\Delta\Om{S}{\dyn}$}	\\
			\hline
			$^1\Pi_g(n \ra \pis)$			&	Val.	&	9.90	&	9.58	&	-0.32	&	9.92	&	9.53	&	-0.40	&	10.01	&	9.59	&	-0.42	\\
			$^1\Sigma_u^-(\pi \ra \pis)$	&	Val.	&	9.70	&	9.37	&	-0.33	&	9.61	&	9.19	&	-0.42	&	9.69	&	9.25	&	-0.44	\\
			$^1\Delta_u(\pi \ra \pis)$		&	Val.	&	10.37	&	10.05	&	-0.31	&	10.27	&	9.88	&	-0.39	&	10.34	&	9.93	&	-0.41	\\
			$^1\Sigma_g^+$					&	Ryd.	&	15.67	&	15.50	&	-0.17	&	15.04	&	14.84	&	-0.21	&	14.72	&	14.43	&	-0.21	\\
			$^1\Pi_u$						&	Ryd.	&	15.00	&	14.79	&	-0.21	&	14.75	&	14.48	&	-0.27	&	14.80	&	14.59	&	-0.29	\\
			$^1\Sigma_u^+$					&	Ryd.	&	22.88\fnm[1]	&	22.73	&	-0.15	&	19.03	&	18.95	&	-0.08	&	16.78	&	16.71	&	-0.06	\\  
			$^1\Pi_u$						&	Ryd.	&	23.62\fnm[1]	&	23.51	&	-0.11	&	19.15	&	19.04	&	-0.11	&	16.93	&	16.85	&	-0.09	\\  
			\\
			$^3\Sigma_u^+(\pi \ra \pis)$	&	Val.	&	7.39	&	6.91	&	-0.48	&	7.46	&	6.87	&	-0.59	&	7.59	&	6.97	&	-0.62	\\  
			$^3\Pi_g(n \ra \pis)$			&	Val.	&	8.07	&	7.65	&	-0.42	&	8.14	&	7.62	&	-0.52	&	8.24	&	7.70	&	-0.54	\\  
			$^3\Delta_u(\pi \ra \pis)$		&	Val.	&	8.56	&	8.15	&	-0.41	&	8.52	&	8.00	&	-0.52	&	8.62	&	8.07	&	-0.55	\\  
			$^3\Sigma_u^-(\pi \ra \pis)$	&	Val.	&	9.70	&	9.37	&	-0.33	&	9.61	&	9.19	&	-0.42	&	9.69	&	9.25	&	-0.44	\\
			\hline
			\\
						&					&	\mc{3}{c}{aug-cc-pVDZ ($\Eg^{\GW} = 19.49$ eV)}	
											&	\mc{3}{c}{aug-cc-pVTZ ($\Eg^{\GW} = 19.20$ eV)}
											&	\mc{3}{c}{aug-cc-pVQZ ($\Eg^{\GW} = 19.00$ eV)}		\\
											\cline{3-5} \cline{6-8} \cline{9-11}
			State		&	Nature			&	\tabc{$\Om{S}{\stat}$}	&	\tabc{$\Om{S}{\dyn}$}	&	\tabc{$\Delta\Om{S}{\dyn}$} 	
											&	\tabc{$\Om{S}{\stat}$}	&	\tabc{$\Om{S}{\dyn}$}	&	\tabc{$\Delta\Om{S}{\dyn}$}	
											&	\tabc{$\Om{S}{\stat}$}	&	\tabc{$\Om{S}{\dyn}$}	&	\tabc{$\Delta\Om{S}{\dyn}$}	\\
			\hline
			$^1\Pi_g(n \ra \pis)$			&	Val.	&	10.18	&	9.77	&	-0.41	&	10.42	&	9.99	&	-0.42	&	10.52	&	10.09	&	-0.43	\\
			$^1\Sigma_u^-(\pi \ra \pis)$	&	Val.	&	9.95	&	9.51	&	-0.44	&	10.11	&	9.66	&	-0.45	&	10.20	&	9.75	&	-0.45	\\
			$^1\Delta_u(\pi \ra \pis)$		&	Val.	&	10.57	&	10.16	&	-0.41	&	10.75	&	10.33	&	-0.42	&	10.85	&	10.42	&	-0.42	\\
			$^1\Sigma_g^+$					&	Ryd.	&	13.72	&	13.68	&	-0.04	&	13.60	&	13.57	&	-0.03	&	13.54	&	13.52	&	-0.02	\\
			$^1\Pi_u$						&	Ryd.	&	14.07	&	14.02	&	-0.05	&	13.98	&	13.94	&	-0.04	&	13.96	&	13.93	&	-0.03	\\
			$^1\Sigma_u^+$					&	Ryd.	&	13.80	&	13.72	&	-0.08	&	13.98	&	13.91	&	-0.07	&	14.08	&	14.03	&	-0.06	\\  
			$^1\Pi_u$						&	Ryd.	&	14.22	&	14.19	&	-0.04	&	14.24	&	14.21	&	-0.03	&	14.26	&	14.23	&	-0.03	\\  
			\\
			$^3\Sigma_u^+(\pi \ra \pis)$	&	Val.	&	7.75	&	7.12	&	-0.63	&	8.02	&	7.38	&	-0.64	&	8.12	&	7.48	&	-0.64	\\  
			$^3\Pi_g(n \ra \pis)$			&	Val.	&	8.42	&	7.88	&	-0.54	&	8.66	&	8.10	&	-0.56	&	8.75	&	8.20	&	-0.56	\\  
			$^3\Delta_u(\pi \ra \pis)$		&	Val.	&	8.86	&	8.32	&	-0.54	&	9.04	&	8.48	&	-0.56	&	9.14	&	8.57	&	-0.56	\\  
			$^3\Sigma_u^-(\pi \ra \pis)$	&	Val.	&	9.95	&	9.51	&	-0.44	&	10.11	&	9.66	&	-0.45	&	10.20	&	9.75	&	-0.45	\\
		\end{tabular} 
	\end{ruledtabular}
		\fnt[1]{Excitation energy larger than the fundamental gap.}		
\end{table*}
\end{squeezetable}

First, we investigate the basis set dependence of the dynamical correction.
The singlet and triplet excitation energies of the nitrogen molecule \ce{N2} computed at the BSE@{\GOWO}@HF level for the cc-pVXZ and aug-cc-pVXZ families of basis sets are reported in Table \ref{tab:N2}, where we also report the $GW$ gap, $\Eg^{\GW}$, to show that corrected transitions are usually well below this gap.
The \ce{N2} molecule is a very convenient example for this kind of study as it contains $n \ra \pis$ and $\pi \ra \pis$ valence excitations as well as Rydberg transitions.
As we shall further illustrate below, the magnitude of the dynamical correction is characteristic of the type of transitions.
One key result of the present investigation is that the dynamical correction is quite basis set insensitive with a maximum variation of $0.03$ eV between aug-cc-pVDZ and aug-cc-pVQZ.
It is only for the smallest basis set (cc-pVDZ) that one can observe significant differences. 
\titou{We note further that due to its unbound LUMO, the $GW$ gap of \ce{N2}, and to a lesser extent its BSE excitation energies, are very sensitive to the presence of diffuse orbitals. 
However, the dynamical correction is again very stable, being insensitive to the presence of diffuse orbitals (at least for the lowest optical excitations).}
We can then safely conclude that the dynamical correction converges rapidly with respect to the size of the one-electron basis set, a triple-$\zeta$ or an augmented double-$\zeta$ basis being enough to obtain near complete basis set limit values.
This is quite a nice feature as it means that one does not need to compute the dynamical correction in a very large basis to get a meaningful estimate of its magnitude.

%%% TABLE I %%%
\begin{squeezetable}
\begin{table*}
	\caption{
	Singlet excitation energies (in eV) for various molecules obtained with the aug-cc-pVTZ basis set computed at various levels of theory.
	CT stands for charge transfer.
	\label{tab:BigTabSi}
	}
	\begin{ruledtabular}
		\begin{tabular}{llcdddddddddd}
						&			&			&	\mc{5}{c}{BSE@{\GOWO}@HF}	&	\mc{5}{c}{Wave function-based methods}	\\ 
													\cline{4-8}	\cline{9-13}
			Mol.		&	State	&	Nature	&	\tabc{$\Eg^{\GW}$}	&	\tabc{$\Om{S}{\stat}$}	&	\tabc{$\Om{S}{\dyn}$}	&	\tabc{$\Delta\Om{S}{\dyn}$}	&	\tabc{$Z_{S}$}	
									&	\tabc{CIS(D)}	&	\tabc{ADC(2)}	&	\tabc{CC2}	&	\tabc{CCSD}	&	\tabc{TBE}	\\
			\hline
			\ce{HCl}	&	$^1\Pi$							&	CT		&	13.43	&	8.30	&	8.19	&	-0.11	&	1.009	
																		&	6.07	&	7.97	&	7.96	&	7.91	&	7.84	\\
			\\
			\ce{H2O}	&	$^1B_1(n \ra 3s)$				&	Ryd.	&	13.58	&	8.09	&	8.00	&	-0.09	&	1.007	
																		&	7.62	&	7.18	&	7.23	&	7.60	&	7.17	\\
						&	$^1A_2(n \ra 3p)$				&	Ryd.	&			&	9.79	&	9.72	&	-0.07	&	1.005	
																		&	9.41	&	8.84	&	8.89	&	9.36	&	8.92	\\
						&	$^1A_1(n \ra 3s)$				&	Ryd.	&			&	10.42	&	10.35	&	-0.07	&	1.006	
																		&	9.99	&	9.52	&	9.58	&	9.96	&	9.52	\\
			\\
			\ce{N2}		&	$^1\Pi_g(n \ra \pis)$			&	Val.	&	19.20	&	10.42	&	9.99	&	-0.42	&	1.031	
																		&	9.66	&	9.48	&	9.44	&	9.41	&	9.34	\\
						&	$^1\Sigma_u^-(\pi \ra \pis)$	&	Val.	&			&	10.11	&	9.66	&	-0.45	&	1.029	
																		&	10.31	&	10.26	&	10.32	&	10.00	&	9.88	\\
						&	$^1\Delta_u(\pi \ra \pis)$		&	Val.	&			&	10.75	&	10.33	&	-0.42	&	1.030	
																		&	10.85	&	10.79	&	10.86	&	10.44	&	10.29	\\
						&	$^1\Sigma_g^+$					&	Ryd.	&			&	13.60	&	13.57	&	-0.03	&	1.003	
																		&	13.67	&	12.99	&	12.83	&	13.15	&	12.98	\\
						&	$^1\Pi_u$						&	Ryd.	&			&	13.98	&	13.94	&	-0.04	&	1.004	
																		&	13.64	&	13.32	&	13.15	&	13.43	&	13.03	\\
						&	$^1\Sigma_u^+$					&	Ryd.	&			&	13.98	&	13.91	&	-0.07	&	1.008	
																		&	13.75	&	13.07	&	12.89	&	13.26	&	13.09	\\
						&	$^1\Pi_u$						&	Ryd.	&			&	14.24	&	14.21	&	-0.03	&	1.002	
																		&	14.52	&	14.00	&	13.96	&	13.67	&	13.46	\\
			\\
			\ce{CO}		&	$^1\Pi(n \ra \pis)$				&	Val.	&	16.46	&	9.54	&	9.19	&	-0.34	&	1.029	
																		&	8.78	&	8.69	&	8.64	&	8.59	&	8.49	\\
						&	$^1\Sigma^-(\pi \ra \pis)$		&	Val.	&			&	10.25	&	9.90	&	-0.35	&	1.023	
																		&	10.13	&	10.03	&	10.30	&	9.99	&	9.92	\\
						&	$^1\Delta(\pi \ra \pis)$		&	Val.	&			&	10.71	&	10.39	&	-0.32	&	1.023	
																		&	10.41	&	10.30	&	10.60	&	10.12	&	10.06	\\
						&	$^1\Sigma^+$					&	Ryd.	&			&	11.88	&	11.85	&	-0.03	&	1.005	
																		&	11.48	&	11.32	&	11.11	&	11.22	&	10.95	\\
						&	$^1\Sigma^+$					&	Ryd.	&			&	12.39	&	12.37	&	-0.02	&	1.003	
																		&	11.71	&	11.83	&	11.63	&	11.75	&	11.52	\\
						&	$^1\Pi$							&	Ryd.	&			&	12.37	&	12.32	&	-0.05	&	1.004	
																		&	12.06	&	12.03	&	11.83	&	11.96	&	11.72	\\
			\\
%			\ce{HNO}	&	$^1A''(n \ra \pis)$				&	Val.	&	11.71	&	2.46	&	1.98	&	-0.48	&	1.035	
%																		&	1.80	&	1.68	&	1.74	&	1.76	&	1.74	\\	
%						&	$^1A'$							&	Ryd.	&			&	7.05	&	7.01	&	-0.04	&	1.003	
%																		&	5.81	&	5.73	&	5.72	&	6.30	&	6.27	\\
%			\\
			\ce{C2H2}	&	$^1\Sigma_{u}^-(\pi \ra \pis)$	&	Val.	&	12.28	&	7.37	&	7.05	&	-0.32	&	1.026			
																		&	7.28	&	7.24	&	7.26	&	7.15	&	7.10	\\
						&	$^1\Delta_{u}(\pi \ra \pis)$	&	Val.	&			&	7.74	&	7.46	&	-0.29	&	1.025		
																		&	7.62	&	7.56	&	7.59	&	7.48	&	7.44	\\
			\\
			\ce{C2H4}	&	$^1B_{3u}(\pi \ra 3s)$			&	Ryd.	&	11.49	&	7.64	&	7.62	&	-0.03	&	1.004	
																		&	7.35	&	7.34	&	7.29	&	7.42	&	7.39	\\
						&	$^1B_{1u}(\pi \ra \pis)$		&	Val.	&			&	8.18	&	8.03	&	-0.15	&	1.022	
																		&	7.95	&	7.91	&	7.92	&	8.02	&	7.93	\\
						&	$^1B_{1g}(\pi \ra 3p)$			&	Ryd.	&			&	8.29	&	8.26	&	-0.03	&	1.003	
																		&	8.01	&	7.99	&	7.95	&	8.08	&	8.08	\\
			\\
			\ce{CH2O}	&	$^1A_2(n \ra \pis)$				&	Val.	&	12.00	&	5.03	&	4.68	&	-0.35	&	1.027	
																		&	4.04	&	3.92	&	4.07	&	4.01	&	3.98	\\
						&	$^1B_2(n \ra 3s)$				&	Ryd.	&			&	7.87	&	7.85	&	-0.02	&	1.001	
																		&	6.64	&	6.50	&	6.56	&	7.23	&	7.23	\\
						&	$^1B_2(n \ra 3p)$				&	Ryd.	&			&	8.76	&	8.72	&	-0.04	&	1.003	
																		&	7.56	&	7.53	&	7.57	&	8.12	&	8.13	\\
						&	$^1A_1(n \ra 3p)$				&	Ryd.	&			&	8.85	&	8.84	&	-0.01	&	1.000	
																		&	8.16	&	7.47	&	7.52	&	8.21	&	8.23	\\
						&	$^1A_2(n \ra 3p)$				&	Ryd.	&			&	8.87	&	8.85	&	-0.02	&	1.002	
																		&	8.04	&	7.99	&	8.04	&	8.65	&	8.67	\\
						&	$^1B_1(\si \ra \pis)$			&	Val.	&			&	10.18	&	9.77	&	-0.42	&	1.032	
																		&	9.38	&	9.17	&	9.32	&	9.28	&	9.22	\\
						&	$^1A_1(\pi \ra \pis)$			&	Val.	&			&	10.05	&	9.81	&	-0.24	&	1.026	
																		&	9.08	&	9.46	&	9.54	&	9.67	&	9.43	\\
			\hline
			MAE			&									&			&			&	0.64	&	0.50	&			&			
																		&	0.43	&	0.24	&	0.25	&	0.15	&	0.00	\\
			MSE			&									&			&			&	0.64	&	0.48	&			&			
																		&	0.14	&	0.02	&	0.03	&	0.14	&	0.00	\\
			RMSE		&									&			&			&	0.70	&	0.58	&			&			
																		&	0.55	&	0.33	&	0.33	&	0.20	&	0.00	\\
			Max($+$)	&									&			&			&	1.08	&	0.91	&			&			
																		&	1.06	&	0.54	&	0.57	&	0.44	&	0.00	\\
			Max($-$)	&									&			&			&	0.20	&	-0.22	&			&			
																		&	-1.77	&	-0.76	&	-0.71	&	-0.02	&	0.00	\\
		\end{tabular} 
	\end{ruledtabular}
\end{table*}
\end{squeezetable}

%%% TABLE II %%%
\begin{squeezetable}
\begin{table*}
	\caption{
	Triplet excitation energies (in eV) for various molecules obtained with the aug-cc-pVTZ basis set computed at various levels of theory.
	\label{tab:BigTabTr}
	}
	\begin{ruledtabular}
		\begin{tabular}{llcdddddddddd}
						&			&			&	\mc{5}{c}{BSE@{\GOWO}@HF}	&	\mc{5}{c}{Wave function-based methods}	\\
													\cline{4-8}	\cline{9-13}
			Mol.		&	State	&	Nature	&	\tabc{$\Eg^{\GW}$}	&	\tabc{$\Om{S}{\stat}$}	&	\tabc{$\Om{S}{\dyn}$}	&	\tabc{$\Delta\Om{S}{\dyn}$}	&	\tabc{$Z_{S}$}	
									&	\tabc{CIS(D)}	&	\tabc{ADC(2)}	&	\tabc{CC2}	&	\tabc{CCSD}	&	\tabc{TBE}	\\
			\hline
			\ce{H2O}	&	$^3B_1(n \ra 3s)$				&	Ryd.	&	13.58	&	7.62	&	7.48	&	-0.14	&	1.009	
																		&	7.25	&	6.86	&	6.91	&	7.20	&	6.92	\\
						&	$^3A_2(n \ra 3p)$				&	Ryd.	&			&	9.61	&	9.50	&	-0.11	&	1.007	
																		&	9.24	&	8.72	&	8.77	&	9.20	&	8.91	\\
						&	$^3A_1(n \ra 3s)$				&	Ryd.	&			&	9.80	&	9.66	&	-0.14	&	1.008	
																		&	9.54	&	9.15	&	9.20	&	9.49	&	9.30	\\
			\\
			\ce{N2}		&	$^3\Sigma_u^+(\pi \ra \pis)$	&	Val.	&	19.20	&	8.02	&	7.38	&	-0.64	&	1.032	
																		&	8.20	&	8.15	&	8.19	&	7.66	&	7.70	\\
						&	$^3\Pi_g(n \ra \pis)$			&	Val.	&			&	8.66	&	8.10	&	-0.56	&	1.031	
																		&	8.33	&	8.20	&	8.19	&	8.09	&	8.01	\\
						&	$^3\Delta_u(\pi \ra \pis)$		&	Val.	&			&	9.04	&	8.48	&	-0.56	&	1.031	
																		&	9.30	&	9.25	&	9.30	&	8.91	&	8.87	\\
						&	$^3\Sigma_u^-(\pi \ra \pis)$	&	Val.	&			&	10.11	&	9.66	&	-0.45	&	1.029	
																		&	10.29	&	10.23	&	10.29	&	9.83	&	9.66	\\
			\\
			\ce{CO}		&	$^3\Pi(n \ra \pis)$				&	Val.	&	16.46	&	6.80	&	6.25	&	-0.55	&	1.031	
																		&	6.51	&	6.45	&	6.42	&	6.36	&	6.28	\\
						&	$^3\Sigma^+(\pi \ra \pis)$		&	Val.	&			&	8.56	&	8.06	&	-0.50	&	1.025	
																		&	8.63	&	8.54	&	8.72	&	8.34	&	8.45	\\
						&	$^3\Delta(\pi \ra \pis)$		&	Val.	&			&	9.39	&	8.96	&	-0.43	&	1.024	
																		&	9.44	&	9.33	&	9.56	&	9.23	&	9.27	\\
						&	$^3\Sigma_u^-(\pi \ra \pis)$	&	Val.	&			&	10.25	&	9.90	&	-0.35	&	1.023	
																		&	10.10	&	10.01	&	10.27	&	9.81	&	9.80	\\
						&	$^3\Sigma_u^+$					&	Ryd.	&			&	11.17	&	11.07	&	-0.10	&	1.008	
																		&	10.98	&	10.83	&	10.60	&	10.71	&	10.47	\\
			\\
%			\ce{HNO}	&	$^3A''(n \ra \pis)$				&	Val.	&	11.71	&	1.27	&	0.67	&	-0.60	&	1.036	
%																		&	0.91	&	0.78	&	0.84	&	0.85	&	0.88	\\
%						&	$^3A'(\pi \ra \pis)$			&	Val.	&			&	5.55	&	4.87	&	-0.69	&	1.037	
%																		&	5.72	&	5.46	&	5.44	&	5.49	&	5.61	\\
%			\\
			\ce{C2H2}	&	$^3\Sigma_{u}^+(\pi \ra \pis)$	&	Val.	&	12.28	&	5.83	&	5.32	&	-0.51	&	1.031	
																		&	5.79	&	5.75	&	5.76	&	5.45	&	5.53	\\
						&	$^3\Delta_{u}(\pi \ra \pis)$	&	Val.	&			&	6.64	&	6.23	&	-0.41	&	1.028	
																		&	6.62	&	6.57	&	6.60	&	6.41	&	6.40	\\
						&	$^3\Sigma_{u}^-(\pi \ra \pis)$	&	Val.	&			&	7.37	&	7.05	&	-0.32	&	1.026	
																		&	7.31	&	7.27	&	7.29	&	7.12	&	7.08	\\
			\\
			\ce{C2H4}	&	$^3B_{1u}(\pi \ra \pis)$		&	Val.	&	11.49	&	4.95	&	4.49	&	-0.46	&	1.032	
																		&	4.62	&	4.59	&	4.59	&	4.46	&	4.54	\\
						&	$^3B_{3u}(\pi \ra 3s)$			&	Ryd.	&			&	7.46	&	7.42	&	-0.04	&	1.004	
																		&	7.26	&	7.23	&	7.19	&	7.29	&	7.23	\\
						&	$^3B_{1g}(\pi \ra 3p)$			&	Ryd.	&			&	8.23	&	8.19	&	-0.04	&	1.004	
																		&	7.97	&	7.95	&	7.91	&	8.03	&	7.98	\\
			\\
			\ce{CH2O}	&	$^3A_2(n \ra \pis)$				&	Val.	&	12.00	&	4.28	&	3.87	&	-0.40	&	1.027	
																		&	3.58	&	3.46	&	3.59	&	3.56	&	3.58	\\
						&	$^3A_1(\pi \ra \pis)$			&	Val.	&			&	6.31	&	5.75	&	-0.56	&	1.033	
																		&	6.27	&	6.20	&	6.30	&	5.97	&	6.06	\\
						&	$^3B_2(n \ra 3s)$				&	Ryd.	&			&	7.60	&	7.56	&	-0.05	&	1.002	
																		&	6.66	&	6.39	&	6.44	&	7.08	&	7.06	\\
			\hline
			MAE			&									&			&			&	0.41	&	0.27	&			&			
																		&	0.27	&	0.21	&	0.24	&	0.10	&	0.00	\\
			MSE			&									&			&			&	0.41	&	0.06	&			&			
																		&	0.23	&	0.10	&	0.14	&	0.05	&	0.00	\\
			RMSE		&									&			&			&	0.45	&	0.33	&			&			
																		&	0.31	&	0.27	&	0.30	&	0.13	&	0.00	\\
			Max($+$)	&									&			&			&	0.70	&	0.60	&			&			
																		&	0.63	&	0.57	&	0.63	&	0.29	&	0.00	\\
			Max($-$)	&									&			&			&	0.11	&	-0.39	&			&			
																		&	-0.40	&	-0.67	&	-0.62	&	-0.11	&	0.00	\\
		\end{tabular} 
	\end{ruledtabular}
\end{table*}
\end{squeezetable}

Tables \ref{tab:BigTabSi} and \ref{tab:BigTabTr} report, respectively, singlet and triplet excitation energies for various molecules computed at the BSE@{\GOWO}@HF level and with the aug-cc-pVTZ basis set.
For comparative purposes, excitation energies obtained with the same basis set and several second-order wave function methods [CIS(D), ADC(2), CC2, and CCSD] are also reported.
The highly-accurate TBEs of Refs.~\onlinecite{Loos_2018a,Loos_2019,Loos_2020b} (computed in the same basis) will serve us as reference, and statistical quantities [MAE, MSE, RMSE, Max($+$), and Max($-$)] are defined with respect to these references.
For each excitation, we report the static and dynamic excitation energies, $\Om{S}{\stat}$ and $\Om{S}{\dyn}$, as well as the value of the renormalization factor $Z_S$ defined in Eq.~\eqref{eq:Z}.
As one can see in Tables \ref{tab:BigTabSi} and \ref{tab:BigTabTr}, the value of $Z_S$ is always quite close to unity which shows that the perturbative expansion behaves nicely, and that a first-order correction is probably quite a good estimate of the non-perturbative result.
Moreover, we have observed that an iterative, self-consistent resolution [where the dynamically-corrected excitation energies are re-injected in Eq.~\eqref{eq:Om1}] yields basically the same results as its (cheaper) renormalized version.
Note that, unlike in $GW$ where the renormalization factor lies in between $0$ and $1$, the dynamical BSE renormalization factor $Z_S$ defined in Eq.~\eqref{eq:Z} can be smaller or greater than unity. 
A clear general trend is the consistent red shift of the static BSE excitation energies induced by the dynamical correction, as anticipated in Sec.~\ref{sec:dynW}.

%%% FIG I %%%
\begin{figure*}
	\includegraphics[width=\linewidth]{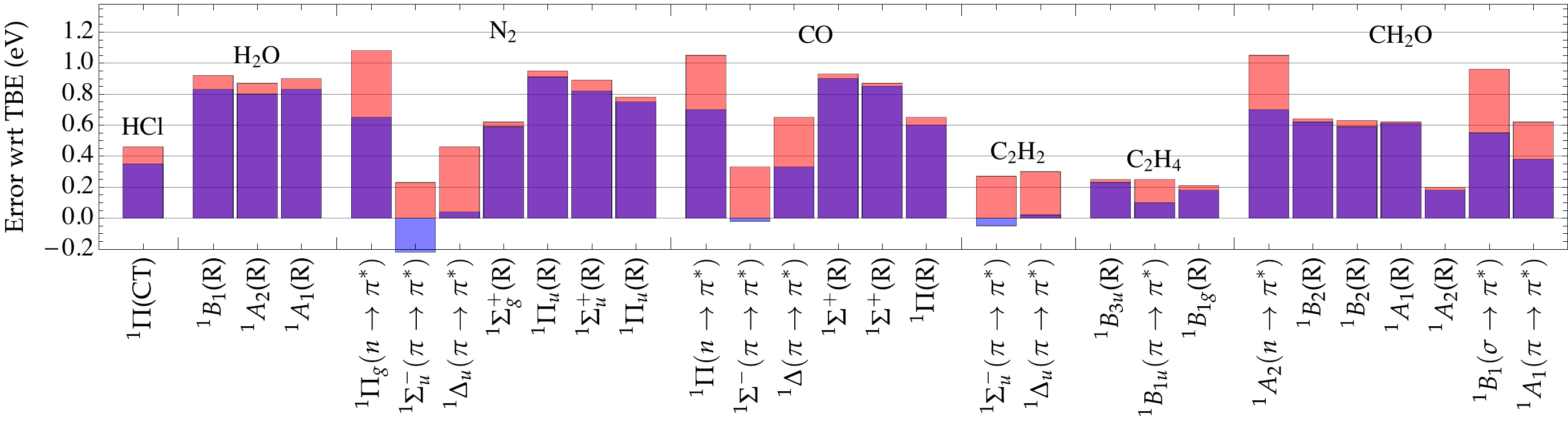}
	\includegraphics[width=\linewidth]{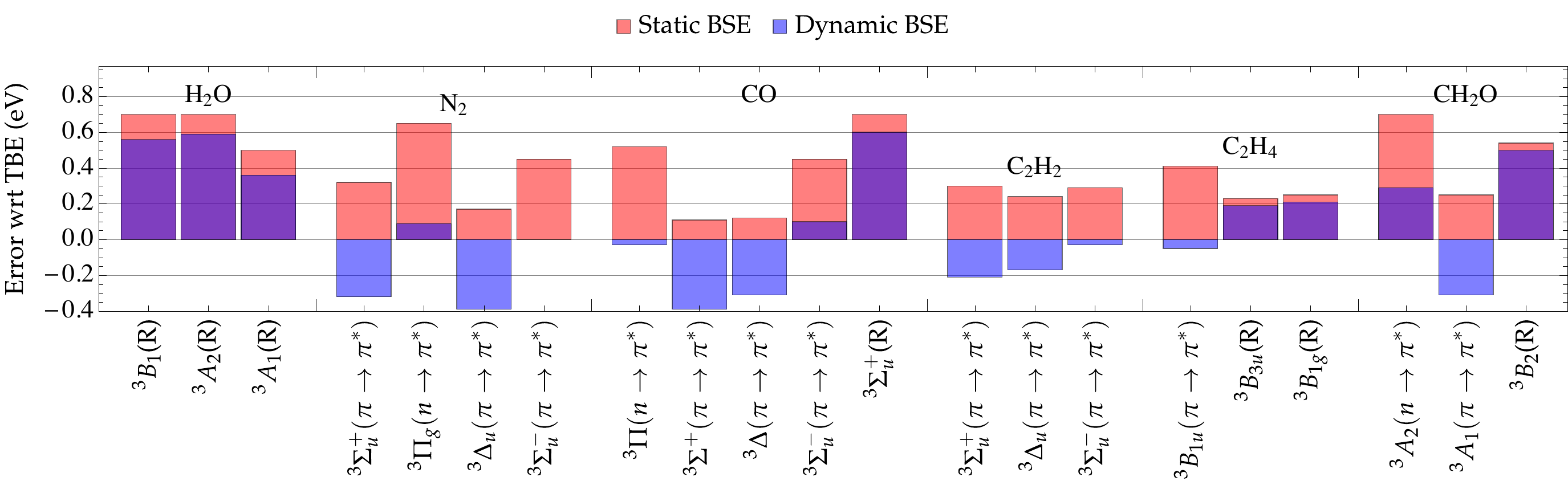}
	\caption{Error (in eV) with respect to the TBEs of Refs.~\onlinecite{Loos_2018a,Loos_2019,Loos_2020b} for singlet (top) and triplet (bottom) excitation energies of various molecules obtained with the aug-cc-pVTZ basis set computed within the static (red) and dynamic (blue) BSE formalism. 
	CT and R stand for charge transfer and Rydberg state, respectively.
	See Tables \ref{tab:BigTabSi} and \ref{tab:BigTabTr} for raw data.
	\label{fig:SiTr-SmallMol}}
\end{figure*}

The results gathered in Tables \ref{tab:BigTabSi} and \ref{tab:BigTabTr} are depicted in Fig.~\ref{fig:SiTr-SmallMol}, where we report the error (with respect to the TBEs) for the singlet and triplet excitation energies computed within the static and dynamic BSE formalism.
From this figure, it is quite clear that the dynamically-corrected excitation energies are systematically improved upon their static analogs, especially for singlet states.
(In the case of triplets, one would notice a few cases where the excitation energies is underestimated.)
In particular, the MAE is reduced from $0.64$ to $0.50$ eV for singlets, and from $0.41$ to $0.27$ eV for triplets.
The MSE and RMSE are also systematically improved when one takes into account dynamical effects.
The second important observation extracted from Fig.~\ref{fig:SiTr-SmallMol} is that the (singlet and triplet) Rydberg states are rather unaltered by the dynamical effects with a correction of few hundredths of eV in most cases.
The same comment applies to the CT excited state of \ce{HCl}.
The magnitude of the dynamical correction for $n \ra \pis$ and $\pi \ra \pis$ transitions is much more important: $0.3$--$0.5$ eV for singlets and $0.3$--$0.7$ eV for triplets.

Dynamical BSE does not quite reach the accuracy of second-order methods [CIS(D), ADC(2), CC2, and CCSD] for the singlet and triplet optical excitations of these small molecules.
However, it is definitely an improvement in terms of performances as compared to static BSE, especially for triplet states, where dynamical BSE reaches an accuracy close to CIS(D), ADC(2), and CC2.

%%% TABLE III %%%
\begin{squeezetable}
\begin{table}
	\caption{
	Singlet and triplet excitation energies (in eV) for various molecules obtained with the aug-cc-pVDZ basis set computed at various levels of theory.
	\label{tab:BigMol}
	}
	\begin{ruledtabular}
		\begin{tabular}{llcdddddd}
						&		&	&	\mc{5}{c}{BSE@{\GOWO}@HF}	\\ 
													\cline{4-8}	
			Molecule			&	State		&	Nature	&	\tabc{$\Eg^{\GW}$}	&	\tabc{$\Om{S}{\stat}$}	&	\tabc{$\Om{S}{\dyn}$}	&	\tabc{$\Delta\Om{S}{\dyn}$}	&	\tabc{$Z_{S}$}	& \tabc{CC3}		\\
			\hline
			acrolein			&	$^1A''(n \ra \pis)$		&	Val.	&	11.67	&	4.62	&	4.28	&	-0.35	&	1.030	&	3.77		\\
								&	$^1A'(n \ra \pis)$		&	Val.	&			&	6.86	&	6.70	&	-0.16	&	1.023	&	6.67		\\
								&	$^1A'(n \ra 3s)$		&	Ryd.	&			&	7.57	&	7.53	&	-0.04	&	1.004	&	6.99		\\
			\\
								&	$^3A''(n \ra \pis)$		&	Val.	&			&	3.97		&	3.54	&	-0.43	&	1.031	&	3.47		\\
								&	$^3A'(\pi \ra \pis)$	&	Val.	&			&	4.03		&	3.61	&	-0.42	&	1.032	&	3.95		\\
			\\
			butadiene			&	$^1B_u(\pi \ra \pis)$	&	Val.	&	9.88	&	6.25	&	6.13	&	-0.12	&	1.019	&	6.25		\\
								&	$^1A_g(\pi \ra \pis)$	&	Val.	&			&	6.88	&	6.86	&	-0.03	&	1.003	&	6.68		\\
			\\
								&	$^3B_u(\pi \ra \pis)$	&	Val.	&			&	3.68	&	3.25	&	-0.43	&	1.032	&	3.36		\\
								&	$^3A_g(\pi \ra \pis)$	&	Val.	&			&	5.51	&	5.01	&	-0.50	&	1.040	&	5.21		\\
								&	$^3B_g(\pi \ra 3s)$		&	Ryd.	&			&	6.29	&	6.25	&	-0.04	&	1.005	&	6.20		\\
			\\
			diacetylene			&	$^1\Sigma_u^-(\pi \ra \pis)$	&	Val.	&	11.01	&	5.62	&	5.35	&	-0.28	&	1.025	&	5.44		\\
								&	$^1\Delta_u(\pi \ra \pis)$		&	Val.	&			&	5.87	&	5.63	&	-0.25	&	1.024	&	5.69		\\
			\\
								&	$^3\Sigma_u^+(\pi \ra \pis)$	&	Val.	&			&	4.30	&	3.82	&	-0.49	&	1.031	&	4.06		\\
								&	$^3\Delta_u(\pi \ra \pis)$		&	Val.	&			&	5.04	&	4.68	&	-0.36	&	1.027	&	4.86		\\
			\\
			glyoxal				&	$^1A_u(n \ra \pis)$		&	Val.		&	10.90	&	3.46	&	3.14	&	-0.33	&	1.028	&	2.90		\\
								&	$^1B_g(n \ra \pis)$		&	Val.		&			&	4.96	&	4.55	&	-0.41	&	1.034	&	4.30		\\
								&	$^1B_u(n \ra 3p)$		&	Ryd.		&			&	7.90	&	7.86	&	-0.04	&	1.004	&	7.55		\\
			\\
								&	$^3A_u(n \ra \pis)$		&	Val.		&			&	2.77	&	2.38	&	-0.39	&	1.028	&	2.49		\\
								&	$^3B_g(n \ra \pis)$		&	Val.		&			&	4.23	&	3.75	&	-0.48	&	1.034	&	3.91		\\
								&	$^3B_u(\pi \ra \pis)$	&	Val.		&			&	5.01	&	4.47	&	-0.55	&	1.034	&	5.20		\\
			\\
			streptocyanine		&	$^1B_2(\pi \ra \pis)$	&	Val.		&	13.79	&	7.66	&	7.51	&	-0.15	&	1.019	&	7.14	\\
			\hline
			MAE					&							&				&			&	0.32	&	0.23	&			&			&	0.00		\\
			MSE					&							&				&			&	0.30	&	0.00	&			&			&	0.00		\\
			RMSE				&							&				&			&	0.38	&	0.29	&			&			&	0.00		\\
			Max($+$)			&							&				&			&	0.85	&	0.54	&			&			&	0.00		\\
			Max($-$)			&							&				&			&	-0.19	&	-0.73	&			&			&	0.00		\\
		\end{tabular} 
	\end{ruledtabular}
\end{table}
\end{squeezetable}

%%% FIG II %%%
\begin{figure*}
	\includegraphics[width=\linewidth]{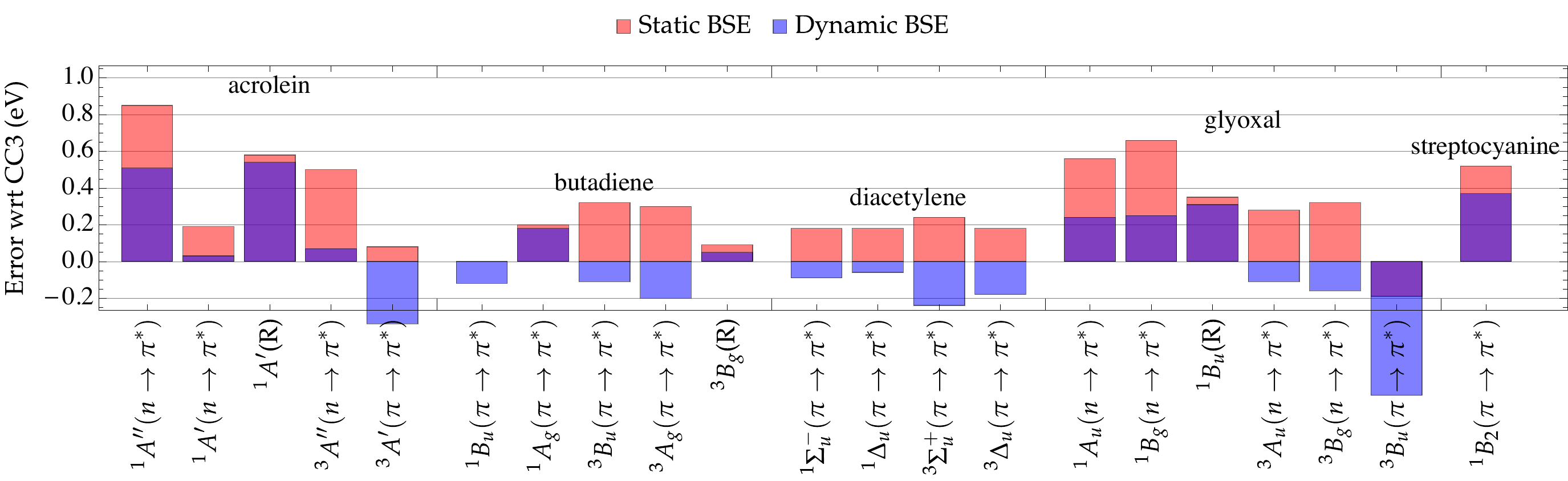}
	\caption{Error (in eV) with respect to CC3 for singlet and triplet excitation energies of various molecules obtained with the aug-cc-pVDZ basis set computed within the static (red) and dynamic (blue) BSE formalism. 
	R stands for Rydberg state.
	See Table \ref{tab:BigMol} for raw data.
	\label{fig:SiTr-BigMol}}
\end{figure*}

Table \ref{tab:BigMol} reports singlet and triplet excitation energies for larger molecules (acrolein \ce{H2C=CH-CH=O}, butadiene \ce{H2C=CH-CH=CH2}, diacetylene \ce{HC#C-C#CH}, glyoxal \ce{O=CH-CH=O}, and streptocyanine-C1 \ce{H2N-CH=NH2+}) at the static and dynamic BSE levels with the aug-cc-pVDZ basis set.
We also report the CC3 excitation energies computed in Refs.~\onlinecite{Loos_2018a,Loos_2019,Loos_2020b} with the same basis set.
These will be our reference as they are known to be extremely accurate ($0.03$--$0.04$ eV from the TBEs). \cite{Loos_2018a,Loos_2019,Loos_2020b,Loos_2020g}
Errors associated with these excitation energies (with respect to CC3) are represented in Fig.~\ref{fig:SiTr-BigMol}.
As expected the static BSE excitation energies are much more accurate for these larger molecules with a MAE of $0.32$ eV, a MSE of $0.30$ eV, and a RMSE of $0.38$ eV.
Here again, the dynamical correction improves the accuracy of BSE by lowering the MAE, MSE, and RMSE to $0.23$, $0.00$, and $0.29$ eV, respectively.
Rydberg states are again very slightly affected by dynamical effects, while the dynamical corrections associated with the $n \ra \pis$ and $\pi \ra \pis$ transitions are much larger and of the same magnitude ($0.3$--$0.6$ eV) for both types of transitions. 
This latter observation is somehow different from the outcomes reached by Rohlfing and coworkers in previous works \cite{Ma_2009a,Ma_2009b} (see Sec.~\ref{sec:intro}) where they observed i) smaller corrections, and ii) that $n \ra \pis$ transitions are more affected by the dynamical screening than $\pi \ra \pis$ transitions. 
The larger size of the molecules considered in Refs.~\onlinecite{Ma_2009a,Ma_2009b} may play a role on the magnitude of the corrections, even though we do not observe here a significant reduction going from small systems (\ce{N2}, \ce{CO}, \ldots) to larger ones (acrolein, butadiene, \ldots). 
We emphasize further that previous calculations \cite{Ma_2009a,Ma_2009b} were performed within the plasmon-pole approximation for modeling the dynamical behaviour of the screened Coulomb potential, while we go beyond this approximation in the present study [see Eq.~\eqref{eq:wtilde}]. 
Finally, while errors were defined with respect to experimental data in Refs.~\onlinecite{Ma_2009a,Ma_2009b}, we consider here as reference high-level CC calculations performed with the very same geometries and basis sets than our BSE calculations. 
As pointed out in previous works, \cite{Loos_2018,Loos_2019b,Loos_2020g} a direct comparison between theoretical transition energies and experimental data is a delicate task, as many factors (such as zero-point vibrational energies and geometrical relaxation) must be taken into account for fair comparisons.
Further investigations are required to better evaluate the impact of these considerations on the influence of dynamical screening.

To provide further insight into the magnitude of the dynamical correction to valence, Rydberg, and CT excitations, let us consider a simple two-level systems with $i = j = h$ and $a = b = l$, where $(h,l)$ stand for HOMO and LUMO. 
The dynamical correction associated with the HOMO-LUMO transition reads
\begin{equation}
	\W{hh,ll}{\text{stat}} - \widetilde{W}_{hh,ll}( \Om{1}{} )
	=     - 4 \sERI{hh}{hl} \sERI{ll}{hl} \qty( \frac{1}{\OmRPA{hl}} - \frac{1}{\Om{hl}{1} - \OmRPA{hl}} ),
\end{equation}
where the only RPA excitation energy, $\OmRPA{hl} = \e{l} - \e{h} + 2 \ERI{hl}{lh}$, is again the HOMO-LUMO transition, \ie, $m=hl$ [see Eq.~\eqref{eq:sERI}]. 
For CT excitations with vanishing HOMO-LUMO overlap [\ie, $\ERI{h}{l} \approx 0$] \titou{and small excitonic binding energy}, $\sERI{hh}{hl} \approx 0$ and $\sERI{ll}{hl} \approx 0$, so that one can expect the dynamical correction to be weak. 
Likewise, Rydberg transitions which are characterized by a delocalized LUMO state, that is, a small HOMO-LUMO overlap, are expected to undergo weak dynamical corrections.  
The discussion for $\pi \ra \pis$ and $n \ra \pis$ transitions is certainly more complex and molecule-specific symmetry arguments must be invoked to understand the magnitude of the $\sERI{hh}{hl}$ and $\sERI{ll}{hl}$ terms. 

As a final comment, let us discuss the two singlet states of butadiene reported in Table \ref{tab:BigMol}.\cite{Maitra_2004,Cave_2004,Saha_2006,Watson_2012,Shu_2017,Barca_2018a,Barca_2018b,Loos_2019}
As discussed in Sec.~\ref{sec:intro}, these corresponds to a bright state of $^1B_u$ symmetry with a clear single-excitation character, and a dark $^1A_g$ state including a substantial fraction of double excitation character (roughly $30\%$).
Although they are both of $\pi \ra \pis$ nature, they are very slightly altered by dynamical screening with corrections of $-0.12$ and $-0.03$ eV for the $^1B_u$ and $^1A_g$ states, respectively. 
The small correction on the $^1A_g$ state might be explained by its rather diffuse nature (similar to a Rydberg states). \cite{Boggio-Pasqua_2004}

%%%%%%%%%%%%%%%%%%%%%%%%
\section{Conclusion}
\label{sec:conclusion}
%%%%%%%%%%%%%%%%%%%%%%%%
The BSE formalism is quickly gaining momentum in the electronic structure community thanks to its attractive computational scaling with system size and its overall accuracy for modeling single excitations of various natures in large molecular systems. 
It now stands as a genuine cost-effective excited-state method and is regarded as a valuable alternative to the popular TD-DFT method.
However, the vast majority of the BSE calculations are performed within the static approximation in which, in complete analogy with the ubiquitous adiabatic approximation in TD-DFT, the dynamical BSE kernel is replaced by its static limit.
One key consequence of this static approximation is the absence of higher excitations from the BSE optical spectrum.
Following Strinati's footsteps \titou{who originally derived the dynamical BSE equations}, \cite{Strinati_1982,Strinati_1984,Strinati_1988} several groups have explored the BSE formalism beyond the static approximation by retaining (or reviving) the dynamical nature of the screened Coulomb potential \cite{Sottile_2003,Romaniello_2009b,Sangalli_2011} or via a perturbative approach coupled with the plasmon-pole approximation. \cite{Rohlfing_2000,Ma_2009a,Ma_2009b,Baumeier_2012b}

In the present study, we have computed exactly the dynamical screening of the Coulomb interaction within the random-phase approximation, going effectively beyond both the usual static approximation and the plasmon-pole approximation.
\titou{Dynamical corrections have been calculated using a renormalized first-order perturbative correction to the static BSE excitation energies following the work of Rohlfing and coworkers. \cite{Rohlfing_2000,Ma_2009a,Ma_2009b,Baumeier_2012b}
Note that, although the present study goes beyond the static approximation of BSE, we do not recover additional excitations as the perturbative treatment accounts for dynamical effects only on excitations already present in the static limit. 
However, we hope to report results on a genuine dynamical approach in the near future in order to access double excitations within the BSE formalism.}
In order to assess the accuracy of the present scheme, we have reported a significant number of calculations for various molecular systems.
Our calculations have been benchmarked against high-level CC calculations, allowing to clearly evidence the systematic improvements brought by the dynamical correction for both singlet and triplet excited states.
We have found that, although $n \ra \pis$ and $\pi \ra \pis$ transitions are systematically red-shifted by $0.3$--$0.6$ eV thanks to dynamical effects, their magnitude is much smaller for CT and Rydberg states.

%%%%%%%%%%%%%%%%%%%%%%%%
\acknowledgements 
%%%%%%%%%%%%%%%%%%%%%%%%
The authors would like to thank Elisa Rebolini, Pina Romaniello, Arjan Berger, and Julien Toulouse for insightful discussions on dynamical kernels. 
PFL thanks the European Research Council (ERC) under the European Union's Horizon 2020 research and innovation programme (Grant agreement No.~863481) for financial support.
This work was performed using HPC resources from GENCI-TGCC (Grant No.~2019-A0060801738) and CALMIP (Toulouse) under allocation 2020-18005.
Funding from the \textit{``Centre National de la Recherche Scientifique''} is acknowledged.
This study has been (partially) supported through the EUR grant NanoX No.~ANR-17-EURE-0009 in the framework of the \textit{``Programme des Investissements d'Avenir''.}

%%%%%%%%%%%%%%%%%%%%%%%%
\section*{Data availability}
%%%%%%%%%%%%%%%%%%%%%%%%
The data that support the findings of this study are available within the article.

\appendix

\section{Fourier transform of $L_0(1,4; 1',3)$}
\label{app:A}

In this Appendix, we derive Eqs.~\eqref{eq:iL0} to \eqref{eq:iL0bis}.
Combining the Fourier transform (with respect to $t_1$) of $L_0(1,4;1',3)$
\begin{align}  
	[L_0](\bx_1,4;\bx_{1'},3 \; | \; \omega_1 ) 
	= -i \int dt_1 e^{i \omega_1 t_1 } G(1,3)G(4,1'),
\end{align}
(where $t_{1'} = t_1^{+}$) with the inverse Fourier transform of the Green's function, \eg,
\begin{align}
     G(1,3) =  \int \frac{ d\omega }{ 2\pi } G(\bx_1,\bx_3;\omega) e^{-i \omega \tau_{13} },
\end{align}
(where $\tau_{13} = t_1-t_3$), we obtain
\begin{multline}  
	[L_0](\bx_1,4;\bx_{1'},3 \;| \;  \omega_1 )   =
	\\
	\int \frac{ d\omega }{ 2i\pi } \;  G(\bx_1,\bx_3;\omega) \; G(\bx_4,\bx_{1'};\omega-\omega_1)  
	  e^{ i \omega t_3 }	e^{-i (\omega-\omega_1) t_4 }.
\end{multline}
Applying the change of variable $\omega \ra \omega + \omega_1/2$, one gets
\begin{multline}  
	[L_0](\bx_1,4;\bx_{1'},3 \; | \; \omega_1 )     = 
	\\
	e^{ i  \omega_1  t^{34} }
	\int \frac{ d\omega }{ 2i\pi } \;   G\qty(\bx_1,\bx_3;\omega+ \frac{\omega_1}{2} )  G\qty(\bx_4,\bx_{1'};\omega-\frac{\omega_1}{2} )  \;
  e^{ i \omega \tau_{34} }	
\end{multline}
with $\tau_{34} = t_3 - t_4$ and $t^{34}= (t_3+t_4)/2$.
Finally, using the Lehman representation of the Green's functions [see Eq.~\eqref{eq:G-Lehman}], and picking up the poles associated with the occupied (virtual) states in the upper (lower) half-plane for $\tau_{34} > 0$ ($\tau_{34} < 0$), one obtains, using the residue theorem,
\begin{equation} \label{eq:A}
\begin{split}
    & \int \frac{ d \omega }{2i\pi}    \; G\qty(\bx_1,\bx_3; \omega +  \homu  )    G\qty(\bx_4,\bx_{1'}; \omega - \homu  )  e^{  i    \omega  \tau    } 
      \\
         & = \sum_{bj}   
     \frac{ \phi_b(\bx_1) \phi_b^*(\bx_3) \phi_j(\bx_4) \phi_j^*(\bx_{1'})} { \omega_1   - (\e{b} - \e{j}) + i\eta    }  
   \qty[  \theta(\tau) e^{i ( \e{j} + \homu ) \tau }  + \theta(-\tau) e^{i ( \e{b} - \homu ) \tau } ] 
  \\
  & -  \sum_{bj} \frac{ \phi_j(\bx_1) \phi_j^*(\bx_3) \phi_b(\bx_4) \phi_b^*(\bx_{1'})} {  \omega_1   + (\e{b} - \e{j} ) -i\eta } 
     \qty[  \theta(\tau)  e^{i ( \e{j} - \homu ) \tau } + \theta(-\tau)  e^{i ( \e{b} + \homu ) \tau }    ] 
     \\
 & +  \sum_{ab} \text{pp}  + \sum_{ij} \text{hh},
\end{split}
\end{equation}
with $\tau = \tau_{34}$, and where pp and hh label the particle-particle and hole-hole channels (respectively) that are neglected here. \cite{Strinati_1988}
Projecting onto $\phi_a^*(\bx_1) \phi_i(\bx_{1'})$ selects the first line  of the right-hand-side of Eq.~\eqref{eq:A}, yielding Eq.~\eqref{eq:iL0bis}
with $\omega_1 \to \Om{s}{}$.

\section{ $\mel{N}{T [\hpsi(6) \hpsi^{\dagger}(5)] }{N,S}$ in the electron-hole product basis}
\label{app:B}

We now derive in more details Eq.~\eqref{eq:spectral65}. 
Starting with
\begin{equation}
\begin{split}
	\mel{N}{T [\hpsi(6) \hpsi^{\dagger}(5)] }{N,S}
	& = \theta(+\tau_{65}) \mel{N}{ \hpsi(6) \hpsi^{\dagger}(5)  }{N,S} 
	\\
	& - \theta(-\tau_{65}) \mel{N}{  \hpsi^{\dagger}(5) \hpsi(6) }{N,S},
\end{split}
\end{equation}
we employ the relationship between operators in their Heisenberg and Schr\"{o}dinger representations [see Eq.~\eqref{Eisenberg}] to obtain
\begin{equation}
\begin{split}
	& \mel{N}{T [\hpsi(6) \hpsi^{\dagger}(5)]}{N,S} = \\
	& +  \theta(+\tau_{65}) \mel{N}{ \hpsi(\bx_6) e^{-i\hH \tau_{65}} \hpsi^{\dagger}(\bx_5)  }{N,S} e^{ i E^N_0 t_6 } e^{ - i E^N_S t_5 }
	\\
	& - \theta(-\tau_{65}) \mel{N}{  \hpsi^{\dagger}(\bx_5) e^{+ i\hH \tau_{65}} \hpsi(\bx_6) }{N,S} e^{ i E^N_0 t_5 } e^{ - i E^N_S t_6 }.
\end{split}
\end{equation}
Expanding now the field operators with creation/destruction operators in the orbital basis, \ie, 
\begin{align}
    \hpsi(\bx_6)  & = \sum_p \phi_p(\bx_6) \ha_p,
    & 
    \hpsi^{\dagger}(\bx_5)  & = \sum_q \phi_q^{*}(\bx_5) \ha^{\dagger}_q,
\end{align}
one gets
\begin{equation} \label{eq:N65NS}
\begin{split}
     \mel{N}{T [\hpsi(6) \hpsi^{\dagger}(5)]}{N,S}
     \\
     = \sum_{pq} \phi_p(\bx_6)  \phi_q^{*}(\bx_5) 
     [ & \theta(+\tau_{65}) \mel{N}{ \ha_p e^{-i \hH \tau_{65}} \ha^{\dagger}_q  }{N,S} e^{ i E^N_0 t_6 } e^{ - i E^N_S t_5 }  
     \\
      - & \theta(-\tau_{65}) \mel{N}{  \ha^{\dagger}_q e^{+ i \hH \tau_{65}} \ha_p }{N,S} e^{ i E^N_0 t_5 } e^{ - i E^N_S t_6 } ].
\end{split}
\end{equation}
Assuming now that the $\lbrace \e{p} \rbrace$'s are proper addition/removal energies, such as the $GW$ quasiparticle energies, one can use the following relationships
\begin{subequations}
\begin{align}
  e^{+i\hH \tau_{65} } \ha^{\dagger}_p \ket{N} &=
    e^{+i \qty( E^N_0 + \e{p} ) \tau_{65} } \ket{N},
    \\
      e^{ -i\hH \tau_{65} } \ha_q \ket{N} &=
    e^{-i \qty( E^N_0 - \e{q} ) \tau_{65} } \ket{N},
\end{align}
\end{subequations}
that plugged into Eq.~\eqref{eq:N65NS} yield 
\begin{equation}
\begin{split}
	\mel{N}{T [\hpsi(6) \hpsi^{\dagger}(5)]}{N,S} 
	\\
	= \sum_{pq} \phi_p(\bx_6)  \phi_q^{*}(\bx_5) 
    [  & \theta(+ \tau_{65}) \mel{N}{ \ha_p   \ha^{\dagger}_q  }{N,S} e^{ -i \e{p} \tau_{65} } e^{ - i \Om{S}{} t_5 } 
     \\
     - & \theta(-\tau_{65}) \mel{N}{  \ha^{\dagger}_q \ha_p }{N,S} e^{ -i \e{q} \tau_{65}  } e^{ - i \Om{S}{} t_6 } ],
\end{split}
\end{equation}
leading to Eq.~\eqref{eq:spectral65} with $\Om{S}{} = E^N_S - E^N_0$, $t_6 = \tau_{65}/2 + t^{65}$, and $t_5 = - \tau_{65}/2 + t^{65}$.

\end{document}